\documentclass[aps,prd,eqsecnum,twocolumn,epsf]{revtex4}
\usepackage{amsfonts,latexsym,eucal,amsmath,amssymb,times}
\usepackage[dvips]{graphicx}

\newcommand{\vect}[1]{\!\!\!\mbox{ \boldmath $#1$}}

\begin{document}

\thispagestyle{empty}


\title{One-loop Corrections to Scalar and Tensor Perturbations during Inflation
in Stochastic Gravity}

\author{Yuko Urakawa$^{1}$}
\email{yuko_at_gravity.phys.waseda.ac.jp}
\author{Kei-ichi Maeda$^{1\,,2\,,3}$}
\email{maeda_at_waseda.jp}
\address{\,\\ \,\\
$^{1}$ Department of Physics, Waseda University,
Okubo 3-4-1, Shinjuku, Tokyo 169-8555, Japan\\
$^{2}$ Advanced Research Institute for Science and Engineering,
 Waseda University,
Okubo 3-4-1, Shinjuku, Tokyo 169-8555, Japan\\
$^{3}$ Waseda Institute for Astrophysics, Waseda University,
Okubo 3-4-1, Shinjuku, Tokyo 169-8555, Japan}



\preprint{200*-**-**, WU-AP/***/**, hep-th/*******}


\begin{abstract}
Based on the stochastic gravity, we study the loop corrections to the
 scalar and tensor perturbations during inflation. Since the loop
 corrections to scalar perturbations suffer infrared (IR) divergence,
 we consider the IR regularization to obtain the finite value. We find
 that the loop corrections to the scalar perturbations are amplified by
 the e-folding; in other words there appear the logarithmic correction,
 just as discussed by M.Sloth et al. On the other hand, we find that the
 tensor perturbations do not suffer from infrared divergence.
\end{abstract}


\pacs{04.50.+h, 04.70.Bw, 04.70.Dy, 11.25.-w}
\maketitle


\section{Introduction}
Inflation provides a natural framework explaining both the
large-scale homogeneity of the universe and its small-scale irregularity.
Despite its attractive aspects, there are still many unknowns about the
inflation theory, since in most models inflation takes place on an
energy scale many orders of magnitude higher than that accessible by 
accelerators. This is why it is
necessary to learn all we can about this high energy regime
from the signatures left by inflation in the present universe
\cite{Lidsey:1995np, Bassett:2005xm, Lyth:2007qh, Linde:2007fr}.

However, when we consider the power spectrum of the curvature
perturbation $\zeta$ only by linear analysis, many inflation models
predict the same results, which are compatible with the observational data,
although the fundamental theories are quiet different.
To discriminate between different inflationary 
models, it is important to take into account nonlinear 
effects \cite{Bartolo:2004if, Maldacena:2002vr, Seery:2005wm,
Seery:2005gb, Weinberg:2005vy, Weinberg:2006ac, Sloth:2006az,
Sloth:2006nu, Seery:2007we, Seery:2007wf, Lyth:2007jh, Jarnhus:2007ia}.  
In particular, the classical perturbation 
theory predicts that when we consider most inflation models, the
curvature perturbation $\zeta$, which is directly related to the fluctuation of
the temperature of CMB, is conserved in the superhorizon
region \cite{Wands:2000dp, Malik:2003mv, Lyth:2004gb}. 
In this case, the primordial
perturbation is essentially characterized by the behavior of the
background inflaton field near the time of horizon exit. Although this 
fact makes the computation of the generated primordial perturbations
simple, it makes it difficult to discriminate different
inflation models. That is why
the non-local dependence on the evolution of the background scalar
field has been studied among the nonlinear quantum effects such as the loop
corrections \cite{Weinberg:2005vy, Weinberg:2006ac}.  Despite their
importance, it is difficult to compute 
these non-linear quantum effects.
  This is because they contain integrations regarding internal momenta
\cite{Sloth:2006az, Sloth:2006nu, Seery:2007we, Seery:2007wf}.
Furthermore, there are several types of  nonlinear 
interactions that induce loop corrections, such as self-interaction of 
a scalar field and interaction between a matter field and a gravitational
field. Depending on the interaction term, 
we find different loop correction behavior.

Stochastic gravity may be well-suited to computing loop
corrections induced from interactions between a scalar
field and a gravitational field. Stochastic gravity was proposed as a
means of discussing the behavior of the gravitational field on the
sub-Planck scale, which is affected by quantum matter fields
\cite{Hu:2003qn, Hu:1989db, Hu:1999mm, Martin:1999ih, Martin:2000dda,
Hu:2002jm, Hu:2004gf, Roura:1999qr, Roura:2007jj,Hu:2008rg}.
  From our naive expectation, on this energy scale,
the quantum fluctuation of the matter field may dominate that
of the gravitational field. Based on this insight,
Martin and Verdaguer have presented the evolution equation of the
gravitational field, which is affected by a quantum scalar
field \cite{Martin:1999ih}.  The effect induced by the 
quantum matter field is evaluated by the 
so-called closed time path (CPT) formalism 
\cite{Schwinger:1960qe, Chou:1984es, Jordan:1986ug,
Calzetta:1986ey, Su:1987}. 
We integrate the action over quantum scalar fields. As a 
result the evolution equation of the gravitational field is described by
a Langevin-type equation, which is called the Einstein-Langevin
equation.
In general, we need great effort to
compute the loop corrections. In stochastic gravity, 
however, by focusing on non-linear interaction
between a scalar field and the gravitational field which gives the
 leading contribution,
we can compute such loop corrections much more easy.
Hence, in this paper, using the Einstein-Langevin equations
formulated in  \cite{Martin:1999ih}, we evaluate one loop corrections
induced by a quantum scalar field.
 
In our previous work \cite{YU20071}, we applied this formalism to 
the linear perturbations, especially to the 
curvature perturbation $\zeta$, which is important to consider the
imprint on observational data. We find that it
reproduces the same results as the prediction obtained by
 the quantization of the gauge invariant variables
\cite{Mukhanov:1990me, Sasaki:1986hm}, except for the limited case. 
Only when the e-folding from the horizon crossing 
time to the end of inflation exceeds some critical value, does
 the Einstein-Langevin equation 
not give the same result as that of the gauge invariant
variables
\cite{footnote1}. 
Hence, we evaluate loop corrections to scalar and
tensor perturbations 
assuming that  the e-folding is smaller
than the critical value.  

In general, loop corrections contain a divergent part. In a quantum
field theory in Minkowski spacetime, the divergence usually appears
on the high energy scale. To discuss finite and physical
quantities, appropriate
 regularization and renormalization are required. Apart from such an
ultraviolet divergence, there may appear  another
divergence in de Sitter (or quasi de Sitter)
spacetime. 
 This infrared problem is important because if we
 introduce an infrared cut-off to obtain a finite value, 
which gives a logarithmic correction.  Such a logarithmic
 correction amplifies the perturbations.
We also find  that there
 is a crucial difference between the logarithmic corrections
in scalar and tensor perturbations,
which is related to the infrared divergence.

In this paper, 
we consider a minimally coupled single-field inflation as a simple
slow-roll inflation model, whose action is given by
\begin{widetext}
\begin{eqnarray}
 S[g,~\phi] 
 \hspace{0.2cm} =  \int d^4 x \sqrt{-g} \Bigl[ \frac{1}{2 \kappa^2_B}
 ( R - 2 \Lambda_B) + \alpha_B C_{abcd} C^{abcd} +
 \beta_B R^2   
 - \frac{1}{2} \{g^{ab} \partial_a \phi
 \partial_b \phi + 2 V(\phi) \}\Bigr]  \label{total action}
\end{eqnarray}
\end{widetext}
where $\kappa^2_B \equiv 8 \pi G_B$ is the bare gravitational constant. 
The subscript ``B''
represents the values of bare coupling constants. 
After we regularize 
divergent parts and renormalize them, we set the renormalized constants as
$\alpha = \beta = \Lambda=0 $ for simplicity. 
 We also represent 
the renormalized gravitational constant by
$\kappa^2 \equiv 8 \pi G$.
In order to characterize the slow-roll inflation, we adopt two slow-roll
parameters: $\varepsilon \equiv - \dot{H}/H^2$ and
$\eta_V \equiv V_{, \phi \phi}/ \kappa^2 V$. As for the time variable,
we use the conformal time, $\tau$, and represent the time derivative by
a prime.

The paper is organized as follows. In Sec. \ref{Stochastic gravity}, 
we briefly review the basic idea of stochastic 
gravity and consider the properties of  the Einstein-Langevin equation, 
which describes the evolution
of the gravitational field affected by quantum scalar fields. Then we
consider perturbations of the Einstein-Langevin equation around an
inflationary background spacetime. In Sec. \ref{depencence on vertex},
 we discuss how the loop corrections depend on the potential of the
 scalar field in diagrammatical language. 
This part is independent from the computation of the loop
corrections in the later sections. In Sec. \ref{Perturbation of EL eq}, we
consider the perturbation of the Einstein-Langevin equation. Solving
this perturbed equation, we can compute the primordial perturbations
generated from the quantum fluctuation of the scalar field. The quantum  
fluctuation of the scalar field is represented by the stochastic variable.
In Sec. \ref{Noise kernel}, we compute the
correlation function of the stochastic variables. Taking into
account the results in Sec. \ref{Perturbation of EL eq} and
Sec. \ref{Noise kernel}, we evaluate the loop corrections to the
scalar and tensor perturbations. 
The conclusion and discussion follow in Sec. \ref{Discussion}.

\section{Stochastic gravity} 
 \label{Stochastic gravity}
First we shortly summarize
the basic points of stochastic gravity and the Einstein-Langevin
equation derived in \cite{Martin:1999ih}.
It is a generalization of the semi-classical gravity theory.
Assuming that quantum fluctuations of matter fields
dominate that of the gravitational field,
they quantize matter fields but treat gravity as a classical
field. 
The fluctuations of the gravitational field induced through interaction with
quantum matter fields are taken into account as stochastic variables. 
To discuss such gravitational field dynamics,
 the CTP formalism is useful. They derive the
effective equation of motion based on the CTP functional technique
applied to a system-environment interaction, more specifically, 
based on the
influence functional formalism of Feynman and Vernon. It is worth while
noting that this Langevin-type equation is well-suited 
not only to understanding the properties of inflation and the origin of 
large-scale structures in the Universe but also to explaining the 
transition from quantum fluctuations to classical seeds.
In addition to the ordinary Einstein-Hilbert action, this CTP effective
action contains two specific terms, which describe the effects induced
through interaction with quantum matter fields. One is
a memory term, by which the equation of motion depends on the history
of the gravitational field itself. The other is a
stochastic source $\xi_{ab}$, which describes quantum fluctuation of
a scalar field. 
The latter is obtained from the imaginary part of
the effective action, and as such it cannot be interpreted as a 
conventional action.
Indeed, there appear 
statistically weighted  stochastic noises as a source for the
gravitational field. Under the Gaussian approximation, this stochastic
variable is characterized by the average value and the
two-point correlation function: 
\begin{eqnarray}
&&
\langle \xi_{ab}(x) \rangle = 0~,
\nonumber \\
&&
 \langle \xi_{ab}(x_1) \xi_{c'd'} (x_2) \rangle = N_{abc'd'}(x_1,~x_2)
\,,   \label{statistics of xi}
\end{eqnarray}
where
the bi-tensor $N_{abc'd'}(x_1,~x_2)$ is called a noise kernel,
 which represents 
quantum fluctuation of the energy-momentum tensor in a background
spacetime, i.e.,
\begin{eqnarray}
&&
 N_{abc'd'} (x_1, x_2) \equiv 
 \frac{1}{4} \mbox{Re} [F_{abc'd'} (x_1, x_2)]
 \nonumber\\
&&
= \frac{1}{8} \langle \{ \hat{T}_{ab}(x_1) - \langle \hat{T}_{ab}(x_1)
 \rangle, \hat{T}_{ab}(x_2) - \langle \hat{T}_{ab}(x_2) \rangle \} 
 \rangle[g]  \label{def of Noise kernel}
\,,
 \nonumber 
\\
\end{eqnarray}
where $\{\hat{X},\hat{Y}\}=\hat{X} \hat{Y}+ \hat{Y}\hat{X}$,
 $g$ is the metric of a background spacetime, and 
the bi-tensor $F_{abc'd'} (x,y)$ is defined by
\begin{eqnarray}
F_{abc'd'} (x_1, x_2)
&\equiv& \langle \hat{T}_{ab}(x_1) \hat{T}_{c'd'}(x_2) \rangle [g]
 \nonumber \\
&&
 - \langle \hat{T}_{ab} (x_1) \rangle [g] ~
 \langle \hat{T}_{c'd'} (x_2) \rangle [g]  \label{def of F}
\,.
\end{eqnarray}
The expectation value for the quantum scalar field is evaluated in the
background spacetime $g$. 

Including the above-mentioned stochastic source of 
$\xi_{ab}$, the effective equation of motion for the gravitational
field is written as 
\begin{eqnarray}
&&  G^{ab}[g + \delta g]
 = \kappa^2 \left[
 \langle \hat{T}^{ab} \rangle_R [g + \delta g] + 2 \xi^{ab} 
\right]
\,,
\nonumber \\
 \label{EL eq}
\end{eqnarray}
where $\delta g$ is the metric perturbation induced 
by quantum fluctuation of matter fields 
 and stochastic source
 $\xi_{ab}$ is characterized by the average value and the two-point
 correlation function Eq. (\ref{statistics of xi}). 

Note that this equation is the same as the semiclassical Einstein
equation expect for a source term  of stochastic variables $\xi_{ab}$. 
Furthermore,  the expectation value of the energy-momentum tensor
 includes a nonlocal effect as follows.
It consists of three terms as
\begin{eqnarray}
&&
\langle \hat{T}^{ab} \rangle_R [g + \delta g]
 = \langle \hat{T}^{ab}(x) \rangle [g]
  + \langle \hat{T}^{(1)ab}[\phi[g], \delta g](x) \rangle [g]
\nonumber \\
&&
~~
  -2  \int d^4y \sqrt{- g(y)} H^{abcd}[g](x,y) \delta g_{cd}(y) +
  O(\delta g^2) \,,
\nonumber \\
\label{Tabf} 
\end{eqnarray}
where the expectation value of $\hat{T}^{(1)ab}$ and $H^{abcd}$ are
defined below (Eq. (\ref{expectation of T1}) and 
(\ref{def of Dissipation kernel})). 
The evolution equation for a scalar field 
depends on the gravitational field.
As a result, the expectation value of energy-momentum tensor
(Eq. (\ref{Tabf})) depends not only 
directly on the spacetime geometry but also 
indirectly through a scalar field. 
When we perturb a spacetime as
$( g  + \delta g) $, two different changes 
 appear in the
right hand side of Eq. (\ref{Tabf}).
 The second term in Eq. (\ref{Tabf})
represents the direct change, which is described by
 fluctuation of the gravitational field $\delta g$
as
\begin{eqnarray}
 && \langle \hat{T}^{(1)ab}[\phi[g], \delta g](x) \rangle
 \nonumber \\ && \hspace{0.45cm}
 = \Bigl( \frac{1}{2}g^{ab} \delta g_{cd}
  - \delta^a_{~c} g^{be} \delta g_{de}
  - \delta^b_{~c} g^{ae} \delta g_{de} \Bigr)
 \langle \hat{T}^{cd} \rangle [g] ,
 \nonumber \\
&& \hspace{0.45cm} 
- \Bigl\{\Bigl(1 - \frac{2~\varepsilon}{3} \Bigr) \rho
 + \frac{\delta \psi^2}{2 a^2} \Bigr\}
 \Bigl( g^{ac} g^{bd} - \frac{1}{2} g^{ab} g^{cd} \Bigr) \delta g_{cd}
 ,
\nonumber \\ \label{expectation of T1}
\end{eqnarray}
where $\delta \psi^2$ is defined in terms of the quantum fluctuation of the
scalar field $\psi$ as follows:
\begin{eqnarray}
 \delta \psi^2 \equiv \langle \nabla_0 \psi \nabla_0 \psi 
 + \gamma^{ij}  \nabla_i \psi \nabla_j \psi \rangle [g] .
\end{eqnarray}
We can neglect this term safely on the sub-Planck scale because
this term is smaller by the order of $(\kappa H)^2$ than the preceding term.
To derive this expression, 
we have used the background evolution equation for a scalar field. 

The third integral term  in the
r.h.s. of Eq. (\ref{Tabf}) represents the effect from the indirect
change and is characterized by the dissipation kernel, which is given by
\begin{eqnarray}
 &&
H_{abc'd'}(x_1, x_2) = H^{\rm (S)}_{abc'd'} (x_1, x_2) 
+ H^{\rm (A)}_{abc'd'} (x_1, x_2) 
~~~~~~~~~~ \label{def of Dissipation kernel} \\
 &&
~~~~~~
 H^{\rm (S)}_{abc'd'} (x_1, x_2) 
= \frac{1}{4} \mbox{Im} [S_{abc'd'} (x_1, x_2)]
  \label{Spart of Dissipation kernel}
\\
 &&
~~~~~~
 H^{\rm (A)}_{abc'd'} (x_1, x_2) 
= \frac{1}{4} \mbox{Im} [F_{abc'd'} (x_1, x_2)]
\,,
 \label{Apart of Dissipation kernel}  
\end{eqnarray}
where $S_{abc'd'} (x_1, x_2)$ is defined by
\begin{eqnarray}
 S_{abc'd'} (x_1, x_2)
 \equiv \langle T^* \hat{T}_{ab}(x_1) \hat{T}_{c'd'}(x_2) \rangle [g]
 .
\end{eqnarray}
$T^*$ denotes that we take time ordering before we apply the derivative
operators in the energy momentum tensor. As pointed out in
\cite{Martin:1999ih}, only if the background spacetime  
$g$ satisfies the semiclassical Einstein equation, is the
gauge invariance of the Einstein-Langevin equation guaranteed.
Hence, in this paper, to guarantee the gauge invariance, we assume the background
spacetime satisfies the semiclassical Einstein
equation.

The Einstein-Langevin equation
Eq. (\ref{EL eq}) contains two
different sources. One is a stochastic source $\xi_{ab}$, whose
correlation function is given by the noise kernel. From
the explicit form of a noise kernel
Eq. (\ref{def of Noise kernel}), we find that 
$\xi_{ab}$ represents the
quantum fluctuation of the energy momentum tensor. The other is an
expectation value of the energy momentum tensor in the perturbed
spacetime $(g + \delta g)$, which includes a memory term.  The integrand
of a memory term consists of a dissipation kernel and 
fluctuation of the gravitational field. To investigate the
evolution for fluctuation of the gravitational field, it is
necessary to calculate the quantum correction of a scalar field and
evaluate the noise kernel and the dissipation kernel.  
Note that the noise kernel and the dissipation
kernel correspond to the contributions from internal lines or loops of 
the Feynman diagrams,
which consist of propagators of a scalar field
and do not include those of the gravitational field
in our approach.

\section{Generic feature from the vertex operators} 
\label{depencence on vertex}
Before we discuss loop corrections to
the primordial perturbations in detail, we consider the dependence of
loop corrections on a potential of a scalar field 
diagrammatically. 
 
First we divide a scalar field $\phi$ into the classical part
$\phi_{cl}$ and the part of small quantum fluctuation $\psi$. 
Inflation is mainly driven by $\phi_{cl}$. 
Expanding a potential of
the scalar field $V(\phi)$ around $\phi_{cl}$, we can write
\begin{eqnarray}
 V(\phi) &=& V(\phi_{cl}) \Bigl[ 1 + \frac{V'}{V}(\phi_{cl})~\psi
 + \frac{1}{2!} \frac{V''}{V}(\phi_{cl})~\psi^2 +
 \cdot \cdot \cdot \Bigr] \nonumber \\
 &=& V (\phi_{cl}) \displaystyle \sum_{m=0}^{\infty} \frac{1}{m!}
  \alpha^{(m)}~(\kappa \psi)^m ,  \label{expansion of V} 
\end{eqnarray}
where the coefficient $\alpha^{(m)}$ is defined by
\begin{eqnarray}
 \alpha^{(m)} \equiv 
 \frac{d^mV/d\phi^m (\phi_{cl})}{\kappa^m V(\phi_{cl})} 
\,.
\label{def of am}
\end{eqnarray}
Taking into account that during inflation $\phi_{cl}$ changes on the Planck scale, we have normalized $\alpha^{(m)}$ by the
Planck mass. 
 
Similarly, we also perturb the
gravitational field
in the total action. 
Expanding the total action Eq. (\ref{total action}) in terms of the
fluctuation of the gravitational field $\delta g$ and the fluctuation of
the scalar field $\psi$, we find that the following vertices appear:
\begin{eqnarray}
 \alpha^{(1)} \delta g~\psi~,~~ \alpha^{(2)} \delta g~\psi^2~,~~
 \alpha^{(3)} \delta g~\psi^3~,~~\alpha^{(4)} \delta g~\psi^4~
 \cdot \cdot ~~\,.  \label{vertex}
\end{eqnarray}
Although the kinetic term also includes fluctuation terms such as
$\delta g \psi$ and $\delta g \psi^2$, since in
this section we are interested in information about $V(\phi)$, which
is obtained through the coefficients $\alpha^{(m)}$, we do not pay
attention to these terms. Of course, when we evaluate the loop
corrections in the later sections, we take into account all
the fluctuation terms.

\begin{figure}[t]
\begin{center}
\includegraphics[width=7cm]{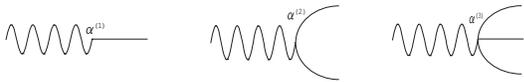}
\caption{Vertices   \label{fg:vertex}}
\end{center}
\end{figure}

The vertex operators given by Eq. (\ref{vertex})
correspond to the vertex diagrams depicted in Fig. \ref{fg:vertex}. 
The solid line represents the propagation of the scalar field
$\psi$. The equation of motion for the field 
in the interaction picture is discussed in
Sec. \ref{Propagator}. As we will see later,
 the propagator in the inflationary universe is proportional to
$H^2$. Hence, each solid line contributes as $(\kappa H)^2$. 
On the other hand, the wavy line
represents the propagation of the gravitational field $\delta g$. The
coefficient $\alpha^{(m)}$ is a coupling constant
of the interaction described by the vertex of $\delta g\, \psi^m$.
Since the higher loop graphs are suppressed 
further by $( \kappa H )^2$, we can discuss the quantum
corrections by an iterative perturbation method.
 
\begin{figure}[h]
\begin{center}
\includegraphics[width=8cm]{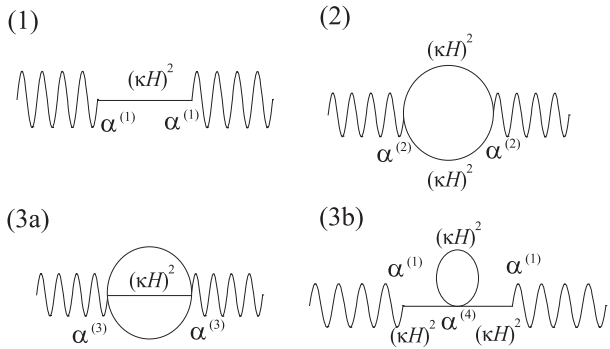}
\caption{Feynman diagrams.}
\label{fig.eps}
\end{center}
\end{figure}

To consider the
evolution equation of the gravitational field, we
integrate out only the degree of freedom of a scalar field.
This means that when we evaluate the effective action in the
CTP formalism, the gravitational field is treated as a
classical external field. 
Hence the gravitational
field contributes as the external line but not as the internal line, 
in the effective action. We represent the gravitational field by the
wavy line. Taking
into account this fact, we find the leading contribution to the effective action
is given by the diagram depicted in Fig. \ref{fig.eps} (1).
The amplitude of this leading diagram is proportional to
$(\alpha^{(1)} \kappa H )^2$; it depends both on the Hubble
parameter $H$ and on the first order derivative of the potential,
$\alpha^{(1)}$. 
The contribution from this tree-level graph corresponds to quantum
corrections in the linear perturbation 
analysis, and it has been given in our previous
work \cite{YU20071}. In the
next leading order depicted by Fig. \ref{fig.eps} (2), the amplitude
 is proportional to
$( \alpha^{(2)})^2 (\kappa H)^4 $. It depends on the
Hubble parameter $H$ and the second order derivative of the potential,
$\alpha^{(2)}$. In this paper, we evaluate
 this leading contribution among loop corrections. 
We just give a few comments on further order contributions.
As seen from Fig. \ref{fig.eps} (3), there are two
different diagrams, whose amplitudes 
are proportional to $(\kappa H)^6$.
Figure \ref{fig.eps} (3a) depends on the third-order derivative of the potential,
$\alpha^{(3)}$, while Fig.\ref{fig.eps} (3b)  depends on the
fourth-order derivative of the potential, $\alpha^{(4)}$.  
Figure \ref{fig.eps} (3b) comes  from the
self-interaction of the scalar field. 
On the other hand, the other three
graphs (Fig. \ref{fig.eps} (1), (2) and (3a)) 
are due to the interaction between the gravitational field and
the scalar field ($\delta g\, \psi^m (m=1,2,3)$). 
From our discussion here, the higher-order
loop corrections, although they are suppressed by the Planck scale,
make it possible to know the more information about the potential of the
scalar field. If we can detect these loop corrections, 
it helps to discriminate many
inflation models, even if they cannot be distinguished only
from the linear perturbation analysis.

\section{Perturbation of Einstein-Langevin equation}
\label{Perturbation of EL eq}
Next we discuss the behavior of the loop corrections,
especially those in the superhorizon region. 
We consider the time evolution of the leading loop corrections,
 which is depicted in Fig. \ref{fig.eps} (2), in the
superhorizon region.
To calculate the loop corrections to the scalar perturbations
and the tensor perturbations, we adopt the following metric form:
\begin{eqnarray}
 ds^2 &=& - a^2(\tau) (1 + 2 \mathcal{A}_{\,\vect{k}} Y_{\,\vect{k}} ) d \tau^2
 - 2 a^2(\tau) \frac{k}{\mathcal{H}}\Phi_{\,\vect{k}} Y_{j \,\vect{k} } d\tau d
 x^j \nonumber \\
 && \hspace{0.5cm}
 +  a^2(\tau)~( \gamma_{ij} + 2 H_{T \,\vect{k}}^{(t)} e_{ij}(\,\vect{k})
 Y_{\,\vect{k}}) dx^i dx^j  \,,
 \label{metric ansatz}
\end{eqnarray}
where $a$ and $\gamma_{ij}$ are the scale factor and the metric of 
maximally symmetric three space. The scalar
perturbations are described by 
$\mathcal{A}$ and $(k/\mathcal{H}) \Phi$, which are
the so-called lapse function and shift vector,
respectively. $H_{T}^{(t)}$ is the tensor perturbation. The scalar
perturbations are expanded by a complete set of harmonic function
$Y_{\,\vect{k}} (\,\vect{x})$, which satisfies
\begin{eqnarray}
 (\Delta + k^2) Y_{\,\vect{k}} (\,\vect{x}) =0
\,.
\end{eqnarray}
Using these harmonic functions, we find the 
scalar components of vector variables are expanded by
\begin{eqnarray}
 Y_{j\,\vect{k}} \equiv - k^{-1} Y_{{\,\vect{k}}|j} \,.
\end{eqnarray} 
The tensor perturbations are expanded by the basis
$e_{ij}({\bf k})$, which satisfies the transverse traceless condition, 
$\gamma^{ij }e_{ij}({\bf k}) = 0$ and $k^i e_{ij}({\bf k}) = 0$. 
Hence our variables are now $ \mathcal{A}_{\,\vect{k}},
\Phi_{\,\vect{k}}$ and $H_{T \,\vect{k}}^{(t)} $.

Since
this gauge choice fixes both the time slicing and the spatial coordinate
completely, all physical variables with this ansatz are gauge
invariant. This choice of the time coordinate is called a flat slicing,
because the spatial curvature vanishes in this slicing.
 
Note that because of  nonlinear perturbations,
the tensor perturbations are not decoupled from the scalar perturbations, and
the quantum fluctuations of the scalar field may amplify not only the
scalar perturbations but also the tensor perturbations.

\subsection{Scalar perturbations}
First we consider  loop corrections to the scalar
perturbations. In particular, we discuss the evolution of the gauge-invariant
variable $\zeta$. We 
focus on proper nonlinear effects, and then
we neglect the contributions
from the product of the linear perturbations. 
Then $\zeta$ is related
to the density perturbation in a flat slicing 
($\delta_f\equiv \delta\rho/\rho$) as 
\begin{eqnarray}
 \zeta = \frac{1}{2 \varepsilon} \delta_f  \label{relation bt zdf}
\,.
\end{eqnarray}
This variable $\zeta$ is gauge-invariant
and  turns out to be a curvature perturbation in
a uniform density slicing. 
In the classical perturbation theory, the
energy conservation law implies that this
variable is conserved
in a superhorizon region for a single-field inflation
\cite{Wands:2000dp,Malik:2003mv,Lyth:2004gb}.
$\zeta$ is directly related
to a gravitational potential at the late stage of the universe
 and then to the observed CMB fluctuations. Hereafter, when we need not
 clarify the mode $~\vect{k}$, we neglect the index of the momentum for
 the perturbed variables.

The density perturbation in the present slicing is given by 
\begin{eqnarray}
 \delta {T}^0_{~0} &\equiv & - \rho \delta_f Y
\nonumber \\
&=&  \delta g_{0c} \langle \hat{T}^{0c}(x) \rangle [g]
 + g_{0c} \{ \langle \hat{T}^{(1)0c}[\phi[g], \delta g](x) \rangle
\nonumber \\
&&   -2  \int d^4y \sqrt{- g(y)} H^{0cde}[g](x,y) \delta g_{de}(y)
 + 2 \xi^{0c} \} \label{relation for df}
\,.
\nonumber \\
\label{density_perturbation}
\end{eqnarray}
Since the
background energy-momentum tensor is given by
\begin{eqnarray}
 \langle \hat{T}^{0a}(x) \rangle [g]
 = g^{0b} \langle \hat{T}^{a}_{~~b} (x) \rangle [g]  = - \rho~g^{0a}
\,,
\end{eqnarray}
the direct contribution from the gravitational
field is described as
\begin{eqnarray}
  \langle \hat{T}^{(1)00}[\phi[g], \delta g](x) \rangle 
 = - \frac{2}{a^2} \Bigl\{ 1 + \frac{\varepsilon}{3} + O((\kappa H)^2) \Bigr\}
 \rho \mathcal{A} Y 
\,.
\nonumber \\
\label{direct_contribution}
\end{eqnarray}
With these two relations (\ref{density_perturbation})  
and (\ref{direct_contribution}), the density perturbation in flat
slicing is written as
\begin{eqnarray}
 \delta_f \simeq - \frac{2 \varepsilon}{3} \mathcal{A}
 + \frac{2}{\rho} ( \delta \rho_m + \delta \rho_{\xi} ) \,,
\label{delta_f}
\end{eqnarray}
where we have defined the density perturbations of the stochastic source
$\xi^a_{~b}$ and of the memory term as follows:
\begin{eqnarray}
 \delta \rho_m 
 &\equiv& \int d^3 \,\vect{x} e^{-i \,\vect{k} \cdot \,\vect{x}}~
 \Bigl[ g^{00} \int d^4y \sqrt{- g(y)} 
\nonumber \\
&&~~~~~~~ 
 \times H_{00c'd'}[g](x,y)
g^{c'e'} g^{d'f'} \delta g_{e'f'}(y) \Bigr] 
\,,
\nonumber \\
  \delta \rho_{\xi} 
&\equiv&
 \int d^3 \,\vect{x} e^{-i \,\vect{k} \cdot \,\vect{x}}~
  \Bigl[- g_{00} ~ \xi^{00}(x) \Bigr]  
\,.
\end{eqnarray}

The Hamiltonian constraint equation gives a relation between the
gauge invariant variable $\mathcal{A}$ and 
the density perturbation
$\delta_f$ as
\begin{eqnarray}
 \mathcal{A} =
 \frac{1}{3} \Bigl( \frac{k}{\mathcal{H}} \Bigr)^2 \Phi 
 - \frac{\delta_f}{2}
\,. \label{Hamiltonian constraint}
\end{eqnarray}
Using it, we eliminate $\mathcal{A}$ in Eq. (\ref{delta_f}),
and find
\begin{eqnarray}
 \Bigl(1 - \frac{\varepsilon}{3} \Bigr) \delta_f = 
 \frac{2}{\rho} (\delta \rho_{\xi} + \delta \rho_m )
 + O \Bigl( (k/ \mathcal{H})^2 \Bigr)  \,.
\end{eqnarray}
Hence, in superhorizon region, the two-point correlation 
function for $\delta_f$ is
expressed in terms of four correlation functions of
$\delta \rho_{\xi}$ and $\delta \rho_m$, i.e.,
\begin{eqnarray}
 && \langle \delta_{f \,\vect{k}}(\tau)
  \delta_{f \,\vect{p}}(\tau) \rangle
 \simeq \frac{4}{V(\tau)^2}  
  \Big[~\langle \delta \rho_{\xi \,\vect{k}}(\tau)
 \delta \rho_{\xi \,\vect{p}}(\tau) \rangle
 \nonumber \\ &&
 \hspace{0.5cm}
 +  \langle \delta \rho_{m \,\vect{k}}(\tau)
 \delta \rho_{\xi \,\vect{p}}(\tau) \rangle
 + \langle \delta \rho_{\xi \,\vect{k}}(\tau)
 \delta \rho_{m \,\vect{p}}(\tau) \rangle
  \nonumber \\ &&
 \hspace{0.5cm}
+ \langle \delta \rho_{m \,\vect{k}}(\tau)
 \delta \rho_{m \,\vect{p}}(\tau) \rangle~\Big]
\,.
 \label{correlation of df} 
\end{eqnarray}
Here we have used the relation
\begin{eqnarray}
V(\tau) = \Bigl(1 - \frac{\varepsilon}{3}\Bigr) \rho +
O(\rho~(H/m_{pl})^2)
\,.
\end{eqnarray}

\subsection{Tensor perturbations}
In a similar way to the scalar perturbations, the transverse traceless
part of the fluctuation of the energy-momentum tensor is given by
\begin{eqnarray}
 \Bigl[ \delta T^i_{~j} \Bigr]_{TT}
 = \Bigl\{ - \frac{\delta \psi^2 }{a^2}~ H_T^{(t)}
 + 2 \Bigl( p \pi_m^{(t)} + p \pi_{\xi}^{(t)} \Bigr) \Bigr\}
 e^i_{~j}(\,\vect{k}) Y_{\,\vect{k}} \,, \nonumber \\
\end{eqnarray}
where we have defined the transverse
traceless part of the anisotropic pressure both for the memory term and
the stochastic variable $\xi^a_{~b}$ as 
\begin{eqnarray}
 p \pi_m^{(t)} e^i_{~j} (\,\vect{k})
 &\equiv& \int d^3 \,\vect{x}  e^{-i \,\vect{k} \cdot \,\vect{x}}~
 \Bigl[- g^{ik}
 \int d^4y \sqrt{- g(y)} \nonumber \\
&&~~~~
 \times H_{kjc'd'}[g](x,y)
 g^{c'e'} g^{d'f'} \delta g_{e'f'}(y) \Bigr]_{TT}
\,,
 \nonumber \\
 p \pi_{\xi}^{(t)} e^i_{~j} (\,\vect{k})
 &\equiv& \int d^3 \,\vect{x} e^{-i \,\vect{k} \cdot \,\vect{x}}~
 [g_{jk} \xi^{ik} (x)]_{TT}
\,.
\end{eqnarray}
Taking into account that the transverse traceless part of Einstein
tensor is written as
\begin{eqnarray}
 \Bigl[\delta G^i_{~j} \Bigr]_{TT} = 
 \frac{1}{a^2} [~\partial_{\tau}^2 + 2 \mathcal{H} \partial_{\tau} + k^2~]
 H_T^{(t)} e^i_{~j}(\,\vect{k}) Y_{\,\vect{k}} \,,
\end{eqnarray}
we find the transverse traceless part of the Einstein-Langevin equation as
\begin{eqnarray}
 && (\partial_{\tau}^2 + 2 \mathcal{H} \partial_{\tau} +  k^2 )
  H_{T~\,\vect{k}}^{(t)}(\tau) \nonumber \\
 &&~~~~= 2 a^2 \kappa^2  \Bigl( p \pi_{m~\,\vect{k}}^{(t)}(\tau)
  + p \pi_{\xi~\,\vect{k}}^{(t)}(\tau) \Bigr) \nonumber \\
 &&~~~~\equiv J_{t \,\vect{k}} (\tau) \,.
 \label{EL eq for tensor}
\end{eqnarray}
The l.h.s. of this equation is the same as the evolution equation
for linear perturbation. In contrast to the linear 
perturbation analysis, where the
tensor perturbations are decoupled from the scalar 
perturbations, the non-linear interaction couples these two modes. That
is why in the r.h.s. there appears the influence of quantum
fluctuations of a scalar field. A linear second-order
differential equation with a source term is solved by the
retarded Green function constructed 
from two independent general solutions. In
the present case, since we are interested in the tensor
perturbations amplified by quantum scalar fields, 
we assume that the tensor
perturbations were absent in the beginning of inflation.
This gives 
$H_{T\,\vect{k}}^{(t)} (\tau_i) = 0$ as the initial condition. The two
independent general solutions for Eq. (\ref{EL eq for tensor}) are given
by 
\begin{eqnarray}
 h_{k}^{(1)} (\tau) = \frac{x^{\frac{1}{2}}}{a(\tau)}  H_{\nu}^{~(1)}(x) 
 ~,~~
 h_k^{(2)} (\tau) = \frac{x^{\frac{1}{2}}}{a(\tau)}
 H_{\nu}^{~(2)}(x) \,, \nonumber \\
\end{eqnarray}
where $\nu^2 = \frac{9}{4} + 3 \varepsilon$ and $x = - k \tau$. Hence,
we find the solution for Eq.(\ref{EL eq for tensor}) as
\begin{eqnarray}
 H_{T\,\vect{k}}^{(t)} (\tau)
 = \int^{\infty}_{\tau_i} d \tau' G_{\rm{ret}~k} (\tau, \tau')
 J_{t~\,\vect{k}} (\tau') \,, \label{general solution for HT}
\end{eqnarray}
where the retarded Green function is given by 
\begin{eqnarray}
 && G_{\rm{ret}\hspace{0.001cm} k}(\tau, \tau')
 = \frac{h^{(1)}_{k}(\tau) h^{(2)}_{k}(\tau')
  - h^{(2)}_{k}(\tau) h^{(1)}_{k}(\tau')}{W_k (\tau')} 
 \theta( \tau - \tau') \nonumber \\
\end{eqnarray}
with
\begin{eqnarray}
&& W_k (\tau) = h^{(2)}_{k}(\tau)~\frac{d~}{d \tau} h^{(1)}_{k}(\tau)
 - h^{(1)}_{k}(\tau)~\frac{d}{d \tau}
 h^{(2)}_{k}(\tau) \,. \nonumber \\
 \label{Wronskian}
\end{eqnarray}
Substituting  general 
solutions into these equations, we obtain 
the corresponding retarded Green function as
\begin{eqnarray}
 && G_{\rm{ret}~k}(\tau, \tau')   \nonumber \\
 &&~~= - \frac{\pi}{2} \frac{a(\tau')}{a(\tau)} 
 \sqrt{\tau \tau'}~{\rm Im}[H_{\nu}^{(1)}(x) H_{\nu'}^{(2)}(x')]~
 \theta( \tau - \tau')  \,. \nonumber \\
\end{eqnarray}
Here we have used the formula for the Hankel functions:
\begin{eqnarray}
 H^{(1)}_{\nu}(x) \frac{d~}{dx} H^{(2)}_{\nu}(x) -
 H^{(2)}_{\nu}(x) \frac{d~}{dx} H^{(1)}_{\nu}(x)=
 \frac{4}{\pi i x} \,.
\end{eqnarray}
Substituting  these expressions into Eq. (\ref{general solution for HT}), 
we find the tensor perturbations amplified by quantum
scalar field as
\begin{eqnarray}
 && H_{T \,\vect{k}}^{(t)} (\tau)   \nonumber \\
 && =  - \frac{\pi}{2} \int^{\tau}_{\tau_i} d \tau'
  \frac{a(\tau')}{a(\tau)}
  \sqrt{\tau \tau'}~{\rm Im} [H_{\nu}^{(1)}(x) H_{\nu}^{(2)}(x')]~
  J_{t \,\vect{k}} (\tau') \nonumber \\
 && = - \frac{\pi}{2 k^2} \int^{1}_{x} d x'
 \Bigl( \frac{x}{x'} \Bigr)^{1 + \varepsilon} 
\sqrt{x x'}~ {\rm Im} [H_{\nu}^{(1)}(x) H_{\nu}^{(2)}(x')]
 J_{t \,\vect{k}} (x') 
 \label{equation for HT}\,.
  \nonumber \\ 
\end{eqnarray}
Here we have used the fact that the scale factor scales as
$a(\tau) \propto |\tau|^{-(1+\varepsilon)}$.
We have also omitted the contribution from the subhorizon region
because the Hankel functions oscillate  where $x$ is larger than one.

\section{Noise kernel}  \label{Noise kernel}
The scalar perturbation $\zeta$ and
the tensor perturbation $H_T^{(t)}$ are given by the
stochastic variable and the memory term. To evaluate the
correlation functions for $\zeta$ and $H_T^{(t)}$, it 
is necessary to compute quantum corrections for the scalar field,
which are imprinted on the noise and dissipation kernels. 
It is expected that the contribution from the memory
terms $\delta \rho_m$ is smaller than that from the stochastic
variable $\delta \rho_\xi$ by the order of magnitude of
the slow-roll parameters.
The reason is as follows. The dissipation kernel is defined as two-point
function of the energy-momentum tensor. As is seen from the definition
of $\alpha^{(m)}$, the contribution from the potential
term is suppressed by the slow-roll parameters. Also, as summarized 
in Appendix B of our paper \cite{YU20071}, the Green function scales as
$(- \tau)^{|\mbox{slow-roll parameter}|}$ in the superhorizon
region. Then, the time derivative of this Green function is suppressed
by the slow-roll parameters. Taking into account that only the contribution
in the superhorizon region can accumulate on the time integral of the
memory term, we can see that the contribution from the memory term,
which is proportional to the dissipation kernel, is suppressed by the
slow-roll parameters. Neglecting the contribution from the memory term, we find 
the density perturbation 
$\delta_f$  only in terms of the density perturbation
of the stochastic variable as
\begin{eqnarray}
 \delta_f \simeq 2 \frac{\delta \rho_{\xi}}{V}  \,. 
 \label{rel bt df rho}
\end{eqnarray}
Similarly, the tensor perturbation $H_T^{(t)}$ is given by the
transverse traceless part of the anisotropic pressure of the stochastic
variable as 
\begin{eqnarray}
 H_{T \,\vect{k}}^{(t)} (\tau) 
  &=& - \frac{\pi \kappa^2}{k^2} \int^{1}_{x} d x'
 \Bigl( \frac{x}{x'} \Bigr)^{1 + \varepsilon} \sqrt{x
 x'} \nonumber \\  &&~~~~\times 
{\rm Im} [H_{\nu}^{(1)}(x) H_{\nu}^{(2)}(x')]~ a^2(\tau')~
 p \pi_{\xi \,\vect{k}}^{(t)}(x')
\,.\nonumber \\ \label{equation for HT2}
\end{eqnarray}
In this section, we shall evaluate the correlation functions of
$\delta \rho_{\xi}$ and $p \pi_{\xi}$. In Appendix 
\ref{computation of N}, we calculate 
these correlation functions  from
the noise kernel. They are expressed in terms of the Wightman Green
function for the interaction picture field in momentum space,
$G_k^+(\tau_1, \tau_2)$. First we determine the Green function, and
then we evaluate the correlation function of $\delta \rho_{\xi}$ and
$p \pi_{\xi}$. 

\subsection{Propagator} \label{Propagator}
As mentioned before, to compute the correlation functions, it is
necessary to determine the Wightman function in momentum space, 
\begin{eqnarray}
 G_k^+ (\tau_1,~\tau_2) \equiv
 \psi_{f, k}(\tau_1) \psi^*_{f, k}(\tau_2) \,,
 \label{Wightman ft in k space}
\end{eqnarray}
where $\psi_{f, k}(\tau)$ is the mode function of
a quantum scalar field, which satisfies the wave equation
\begin{eqnarray}
\psi_{f, k}\hspace{0.001cm}''(\tau)
 + 2 \mathcal{H} \psi_{f, k}\hspace{0.001cm}'(\tau)
 + \{ k^2 + a^2 \kappa^2 V \eta_V  \} \psi_{f, k}(\tau) 
= 0
\,.  \nonumber \\
\label{eq for mode function 0}
\end{eqnarray}
We solve this equation under the slow-roll condition.
Introducing a new variable as
$\tilde{\psi}_{k}(\tau)\equiv a(\tau)~
 \psi_{f, k}(\tau)$,
this equation is rewritten as
\begin{eqnarray}
 \tilde{\psi}_{k}''(\tau) + [k^2 -  
 \{ 2 - \varepsilon - \eta_V ( 3 - \varepsilon) \}
 \mathcal{H}^2 ] \tilde{\psi}_{k}(\tau) = 0  
\,,~~ \label{eq for the mode function 1}
\end{eqnarray}
where we have used the relation
\begin{eqnarray}
  a^2 \kappa^2 V = a^2 \kappa^2 \rho
 \Bigl(1 - \frac{\varepsilon}{3} \Bigr)
 = 3 \mathcal{H}^2 \Bigl(1 - \frac{\varepsilon}{3} \Bigr)
 \,.
\end{eqnarray}

On the sub-Planck scale, we can neglect the term
whose magnitude is smaller 
by the order of $(\kappa H)^2$ than that of the leading term.
We also ignore higher-order terms with respect to
the slow-roll parameters.
So  we do not include the
time evolution of slow-roll parameters. 
Under these assumptions,
the equation for $\tilde{\psi}$ becomes   
\begin{eqnarray}
 \frac{d^2~}{d x^2} \tilde{\psi} (x) + \Bigl[ 1 - \frac{2 + 3
(\varepsilon - \eta_V )}{x^2} \Bigr] \tilde{\psi} (x) = 0
\,,
\end{eqnarray}
where $x \equiv - k \tau $, and we have used
 $\mathcal{H} \simeq -1/[(1 - \varepsilon)\tau]$. 
The general solution
 is given  by the Hankel functions
as
\begin{eqnarray}
 &&  \tilde{\psi}_{k}(\tau) = x^{\frac{1}{2}}
 ~\left[~ \tilde{C} H^{(1)}_{\beta} (x) + \tilde{D} H^{(2)}_{\beta} (x)~
\right] \,,
 \label{def of beta}
\end{eqnarray}
where 
$\beta^2 \equiv {9}/{4} + 3(\varepsilon - \eta_V)$,
with  two arbitrary integration 
constants $\tilde{C}$ and $\tilde{D}$. This implies 
\begin{eqnarray}
 && \psi_{k}(\tau)   = \frac{x^{\frac{1}{2}}}{a(\tau)}
 ~\left[~ \tilde{C} H^{(1)}_{\beta} (x) + \tilde{D} H^{(2)}_{\beta} (x)~
\right] \,.
\end{eqnarray}
We assume that the mode functions
should have the same form as in Minkowski spacetime, i.e.,
\begin{eqnarray}
\psi_{k} (\tau_i) = \frac{1}{\sqrt{2k}} 
~e^{- i k
\tau_i}  \label{IC for psi}
\,,
\end{eqnarray}
when the wavelength is much shorter than the horizon
scale, i.e., at very early stage of the universe.
This fact may be true in the present gauge rather than the comoving gauge.
 Then the mode function and the Wightman function in  momentum
space are given by  
\begin{eqnarray}
 && \psi_{k} (\tau) =
 \frac{\sqrt{\pi |\tau|}}{~2}~
 \frac{a_i}{a(\tau)}~
 e^{i \frac{(2 \beta + 1)\pi}{4}}~H^{(1)}_{\beta} (x) \\
 && G_k^+ (\tau_1,~\tau_2) 
 = \frac{\pi \sqrt{\tau_1~\tau_2}}{~4}~
 \frac{a_i^2}{a_1 a_2}~
 H^{(1)}_{\beta} (x_1)~H^{(2)}_{\beta} (x_2)
\,.
\nonumber \\
\end{eqnarray}
Setting $a_i=1$, we give the scale factor $a(\tau)$ by
$a(\tau) = \left({\tau_i}/{\tau} \right)^{1+ \varepsilon}$. 
Using this fact, we rewrite the Wightman function as
\begin{eqnarray}
 &&
G_k^+ (\tau_1,~\tau_2)
 = \frac{\pi \sqrt{\tau_1~\tau_2}}{~4}
 \Bigl( \frac{\tau_1 \tau_2}{\tau_i^2} \Bigr)^{1 + \varepsilon}
 H^{(1)}_{\beta} (x_1)~H^{(2)}_{\beta} (x_2) 
\nonumber \\
 &&~~=\frac{\pi}{4} \frac{(x_1 x_2)^{\frac{3}{2}}}{k^3}
  \Bigl( \frac{\tau_1 \tau_2}{\tau_i^2} \Bigr)^{\varepsilon}
 (1 - \varepsilon)^2 H_i^2~
 H^{(1)}_{\beta} (x_1)~H^{(2)}_{\beta} (x_2)
\,.
\nonumber \\
\label{G+1}
\end{eqnarray}
Here we have used the relation
\begin{eqnarray}
\tau_i^{-2}= (1- \varepsilon)^2 \mathcal{H}_i^2
 = (1- \varepsilon)^2 H_i^2 \,.
\end{eqnarray}

To compute the correlation functions, it is sufficient to
consider the evolution of the Wightman function in the superhorizon
region. The behavior of $G_k^+ (\tau_1,~\tau_2)$ in the superhorizon region
is summarized in Appendix B in \cite{YU20071}. 

\subsection{Scalar perturbations}
Once the Wightman function is determined, we can compute the correlation
function of $\delta \rho_{\xi}$ from Eq. (\ref{correlation of rho}). 
\begin{eqnarray}
 && \langle \delta \rho_{\xi \,\vect{k}}(\tau)
   \delta \rho_{\xi \,\vect{p}}(\tau)  \rangle^{(4)} \nonumber \\
 && = \int d^3 \,\vect{x}_1  d^3 \,\vect{x}_2
  e^{-i (\,\vect{k} \cdot \,\vect{x}_1+\,\vect{p} \cdot \,\vect{x}_2)}
  \langle \xi^0_{~0}(x_1) \xi^{0'}_{~~0'}(x_2) \rangle^{(4)}
  \Big|_{\tau_1, \tau_2 = \tau}    \nonumber \\
 &&= \frac{1}{8}~\delta (\,\vect{k} + \,\vect{p})
 \int d^3 \,\vect{q} \nonumber \\ 
 &&~~~\times \Bigl\{ (a_1)^{-2}
 \Bigl( \partial^{~~q}_{\tau_1} \partial^{~~k-q}_{\tau_1}
 - \,\vect{q} \cdot (\,\vect{k} - \,\vect{q}) \Bigr)
 + \alpha_1^{(2)} V_1^{cl} \kappa^2 \Bigr\}
 \nonumber \\ && ~~~\times
 \Bigl\{ (a_2)^{-2}
 \Bigl( \partial^{~~q}_{\tau_2} \partial^{~~k-q}_{\tau_2}
 - \,\vect{q} \cdot (\,\vect{k} - \,\vect{q}) \Bigr)
 + \alpha_2^{(2)} V_2^{cl} \kappa^2 \Bigr\}
 \nonumber \\ && ~~~\times
 {\rm Re} \Bigl[ ~G^+_{~q}(\tau_1,~\tau_2)
 G^+_{|\,\vect{k} - \,\vect{q}|}(\tau_1,~\tau_2) \Bigr]
 \Big|_{\tau_1, \tau_2 = \tau}  \,, \nonumber \\
\end{eqnarray}
where the number of the superscript $(4)$ represents the power of $(\kappa H)$.
We put the momentum superscript on the partial derivative operator.
This means, for example, 
$\partial^{~~q}_{\tau_1}$ operates only on the Wightman
function with the momentum $q$, $G^+_{~q}(\tau_1,~\tau_2)$.
It is convenient to divide this correlation function into the
subhorizon part $I_{\rm sb}(\tau,\,\vect{k})$ and the superhorizon part
$I_{\rm sp}(\tau,\,\vect{k})$, which are defined by 
\begin{eqnarray}
 && I_{\rm sb}(\tau,\,\vect{k}) \equiv
 \int_{q \in [\mathcal{H},~ \infty]} d^3 \,\vect{q}
  \nonumber \\ 
 &&~~~\times \Bigl\{ (a_1)^{-2}
 \Bigl( \partial^{~~q}_{\tau_1} \partial^{~~k-q}_{\tau_1}
 - \,\vect{q} \cdot (\,\vect{k} - \,\vect{q}) \Bigr)
 + \alpha_1^{(2)} V_1^{cl} \kappa^2 \Bigr\}
 \nonumber \\ && ~~~\times
 \Bigl\{ (a_2)^{-2}
 \Bigl( \partial^{~~q}_{\tau_2} \partial^{~~k-q}_{\tau_2}
 - \,\vect{q} \cdot (\,\vect{k} - \,\vect{q}) \Bigr)
 + \alpha_2^{(2)} V_2^{cl} \kappa^2 \Bigr\}
 \nonumber \\ && ~~~\times
 {\rm Re} \Bigl[ ~G^+_{~q}(\tau_1,~\tau_2)
 G^+_{|\,\vect{k} - \,\vect{q}|}(\tau_1,~\tau_2) \Bigr]
 \Big|_{\tau_1, \tau_2 = \tau}  \label{Isb} \\
 && I_{\rm sp}(\tau,\,\vect{k}) \equiv
 \int_{q \in [0,~\mathcal{H}]} d^3 \,\vect{q}
   \nonumber \\ 
 &&~~~\times \Bigl\{ (a_1)^{-2}
 \Bigl( \partial^{~~q}_{\tau_1} \partial^{~~k-q}_{\tau_1}
 - \,\vect{q} \cdot (\,\vect{k} - \,\vect{q}) \Bigr)
 + \alpha_1^{(2)} V_1^{cl} \kappa^2 \Bigr\}
 \nonumber \\ && ~~~\times
 \Bigl\{ (a_2)^{-2}
 \Bigl( \partial^{~~q}_{\tau_2} \partial^{~~k-q}_{\tau_2}
 - \,\vect{q} \cdot (\,\vect{k} - \,\vect{q}) \Bigr)
 + \alpha_2^{(2)} V_2^{cl} \kappa^2 \Bigr\}
 \nonumber \\ && ~~~\times
 {\rm Re} \Bigl[ ~G^+_{~q}(\tau_1,~\tau_2)
 G^+_{|\,\vect{k} - \,\vect{q}|}(\tau_1,~\tau_2) \Bigr]
 \Big|_{\tau_1, \tau_2 = \tau} \,.  \label{Isp}
\end{eqnarray}
First we discuss the subhorizon part,
$I_{\rm sb} (\tau,\,\vect{k})$. Since we consider only the superhorizon mode as
the momentum of the external line ($\,\vect{k}$), $k$ is
much smaller than the horizon scale $\mathcal{H}$. This implies that the
momentum of the internal line $\,\vect{q}$ in 
 $I_{\rm sb} (\tau,\,\vect{k})$, which is larger than
$\mathcal{H}$, is much larger than the external momentum, $k$. Hence,
we can approximate $|\,\vect{q} - \,\vect{k}|$ as $q$.
So $I_{\rm sb} (\tau,\,\vect{k})$ depends only on $\tau$,
and then
\begin{eqnarray}
 I_{\rm sb} (\tau,\,\vect{k}) = I_{\rm sb}(\tau) 
  = \frac{1}{k^3} k^3 I_{\rm sb}(\tau)
  \propto \frac{1}{k^3}  (- k \tau)^3
\,.
\end{eqnarray}
Here we have separated the scale invariant part of  $k^{-3}$. 
Since the remaining part, $k^3 I_{\rm sb}(\tau)$, must be a function of 
$- k \tau$, we find that even the leading part of
$I_{\rm sb} (\tau,\,\vect{k})$ decays as $(- k \tau)^3$. Hence we can neglect
the contribution from the subhorizon region.

Consequently, the correlation function of $\delta \rho_{\xi \,\vect{k}}(\tau)$
 is evaluated  only in the superhorizon region as
\begin{eqnarray}
 \langle \delta \rho_{\xi\,\vect{k}}(\tau)
   \delta \rho_{\xi \,\vect{p}}(\tau)  \rangle^{(4)}
 \simeq \frac{1}{8}~\delta (\,\vect{k} + \,\vect{q})~
 I_{\rm sp}(\tau,~ \,\vect{k}) 
\,.
\label{correlation of drho}
\end{eqnarray}
As seen in Appendix B in \cite{YU20071}, in the superhorizon region, the
Wightman function $G^+_k(\tau_1,\tau_2)$ is approximated as
\begin{eqnarray}
  G^{+}_k (\tau_1,~\tau_2) &\simeq& \frac{1}{2}
  \frac{\sqrt{\tau_1 \tau_2}}{a_1 a_2}~
  (x_1 x_2)^{- \beta}  \nonumber \\
  &\simeq& \frac{1}{2}
 \frac{~(\tau_1 \tau_2)^{\frac{1}{2} - \beta}}{a_1 a_2}
 ~k^{- 2 \beta} \,. \label{Re of G+}
\end{eqnarray}
Substituting this expression into the definition of 
$I_{\rm sp}(\tau,\,\vect{k})$ 
and neglecting the sub-leading terms w.r.t. the  
slow-roll parameters, 
we obtain $I_{\rm sp}(\tau,~\,\vect{k})$ as
\begin{eqnarray}
  I_{\rm sp}(\tau,\,\vect{k}) 
   &=& \Bigl(\eta_V V^{cl} \kappa^2 \Bigr)^2
 \frac{1}{4} \frac{~|\tau|^{2-4 \beta}}{a (\tau)^4}
 \nonumber \\
 &\times &~
\int_{q \in [0 ,~\mathcal{H}]} 
 \frac{d^3 \,\vect{q}}{q^{3 + 2 (\varepsilon - \eta_V)}
~{|\,\vect{k} - \,\vect{q}|}^{3 + 2 (\varepsilon - \eta_V)}}
 \,.  \nonumber \\
\end{eqnarray}
Here we encounter the so-called 
infrared divergence problem. In the
long wave limit ($q \rightarrow 0$), the integrand is approximately 
$q^{- [3+  2(\varepsilon - \eta_V)]}$. 
Then this integral could be divergent depending on the signature of
$(\varepsilon - \eta_V)$\cite{footnote2}. 
This is the infrared (IR) problem, which sometimes
appears in the quantum field theory in an inflationary universe. When we
use the scale invariant power spectrum, in general we find
this divergence on the loop corrections.
Here, introducing the cut off by the initial horizon scale, we just neglect
the effects from the long wave modes whose comoving lengths are larger than
the initial horizon scale $a_i H_i$. 
We tentatively discuss this IR problem in Sec.\ref{Discussion} and
elaborate this problem in \cite{YU20081}.

After introducing the cut off $H_i$ and integrating over the internal
momentum $\,\vect{q}$, we obtain a finite result. Using
the loop integral Eq. (\ref{solution for f}), whose detailed derivation is 
given in Appendix C,  we find $I_{\rm sp}(\tau,\,\vect{k})$  as
\begin{eqnarray}
 && I_{\rm sp}(\tau,\,\vect{k})  \nonumber \\
  &&~= \Bigl(\eta_V V^{cl} \kappa^2 \Bigr)^2
 \frac{\pi}{k^3} H(\tau)^4 x^{-4(\varepsilon - \eta_V)}
 \nonumber  \\ &&~\times
 \Bigl[~ \frac{1}{3} - \frac{ 1 - (H_i/k)^{- 2(\varepsilon -
 \eta_V)}}{2(\varepsilon - \eta_V)} 
  - \frac{1}{3} \Bigl( \frac{k}{\mathcal{H}}
 \Bigr)^{3 + 4( \varepsilon - \eta_V )} \Bigr] 
\,.
 \nonumber \\
\end{eqnarray}
Substituting this result into Eq. (\ref{correlation of drho}), we 
obtain the correlation function of $\delta \rho_{\xi \,\vect{k}}(\tau)$ as
\begin{eqnarray}
 && \langle \delta \rho_{\xi \,\vect{k}}(\tau)
   \delta \rho_{\xi \,\vect{p}}(\tau)  \rangle^{(4)} \nonumber \\
 &&~   \simeq \frac{\pi}{8}~
 \frac{\{ \kappa H(\tau) \}^4}{k^3} \Bigl(\eta_V V^{cl} \Bigr)^2~
 x^{4 (\eta_V - \varepsilon)}~
 \delta (\,\vect{k} + \,\vect{p})~
 \nonumber  \\ &&~~\times
 \Bigl[~ \frac{1}{3} - \frac{ 1 - (H_i/k)^{- 2(\varepsilon -
 \eta_V)}}{2(\varepsilon - \eta_V)} 
  - \frac{1}{3} \Bigl( \frac{k}{\mathcal{H}}
 \Bigr)^{3 + 4( \varepsilon - \eta_V )} \Bigr] 
\,.
 \nonumber \\
 \label{correlation of drhof}
\end{eqnarray}
If $\varepsilon > \eta_V $, it
diverges when we remove the cut-off $H_i$.  

\subsection{Tensor perturbations}
Next we calculate the correlation function of the transverse
traceless part of the anisotropic pressure of the stochastic variable,
$p \pi_{\xi \,\vect{k}}(\tau)$, which is given by Eq.(\ref{correlation of ppi})
in Appendix B, as 
\begin{eqnarray}
 && \langle p \pi_{\xi \,\vect{k}}(\tau_1) e^i_{~j} (\,\vect{k})
  p \pi_{\xi \,\vect{p}}(\tau_2) e^{j}_{~i} (\,\vect{p}) \rangle^{(4)}
 \nonumber \\
 &&~= \frac{1}{4 (a_1 a_2)^{2}}~ \delta (\,\vect{k} + \,\vect{p})
 \int d^3 \,\vect{q}~
 \Bigl( q^2 - \frac{(\,\vect{k} \cdot \,\vect{q})^2}{k^2} \Bigr)^2
 \nonumber \\ && \hspace{1cm} \times
  {\rm Re} \Bigl[~G^+_{~q}(\tau_1,~\tau_2)
 G^+_{~|\,\vect{k} - \,\vect{q}|}(\tau_1,~\tau_2)~\Bigr] \,.
  \nonumber \\
\end{eqnarray}
For the tensor perturbations, we also divide the correlation
functions into the subhorizon part $J_{\rm sb}(\tau_1, \tau_2,\,\vect{k})$ and
the superhorizon part $J_{\rm sp}(\tau_1, \tau_2,\,\vect{k})$, which are
defined as
\begin{eqnarray}
 && J_{\rm sb}(\tau_1, \tau_2,\,\vect{k}) \nonumber \\
 &&~~ \equiv
  \theta(\tau_1 - \tau_2 )
   \int_{q \in [\mathcal{H}_2, \infty]} d^3 \,\vect{q}~
   \Bigl( q^2 - \frac{(\,\vect{k} \cdot \,\vect{q})^2}{k^2} \Bigr)^2
   \nonumber \\ && \hspace{2cm} \times
    {\rm Re} \Bigl[~G^+_{~q}(\tau_1,~\tau_2)
    G^+_{~|\,\vect{k} - \,\vect{q}|}(\tau_1,~\tau_2)~\Bigr]
  \nonumber \\ &&~~~+ \theta(\tau_2 - \tau_1 )
  \int_{q \in [\mathcal{H}_1, \infty]} d^3 \,\vect{q}~
   \Bigl( q^2 - \frac{(\,\vect{k} \cdot \,\vect{q})^2}{k^2} \Bigr)^2
   \nonumber \\ && \hspace{2cm} \times
    {\rm Re} \Bigl[~G^+_{~q}(\tau_1,~\tau_2)
    G^+_{~|\,\vect{k} - \,\vect{q}|}(\tau_1,~\tau_2)~\Bigr]
  \nonumber \\ 
\label{Jsb} 
\end{eqnarray}
\begin{eqnarray}
 && J_{\rm sp}(\tau_1, \tau_2,\,\vect{k})
  \nonumber \\ &&~~
 \equiv
 \theta(\tau_1 - \tau_2 )
   \int_{q \in [H_i, \mathcal{H}_2]} d^3 \,\vect{q}~
 \Bigl( q^2 - \frac{(\,\vect{k} \cdot \,\vect{q})^2}{k^2} \Bigr)^2
   \nonumber \\ && \hspace{2cm} \times
    {\rm Re} \Bigl[~G^+_{~q}(\tau_1,~\tau_2)
    G^+_{~|\,\vect{k} - \,\vect{q}|}(\tau_1,~\tau_2)~\Bigr]
  \nonumber \\ &&~~~+ \theta(\tau_2 - \tau_1 )
   \int_{q \in [H_i, \mathcal{H}_1]} d^3 \,\vect{q}~
 \Bigl( q^2 - \frac{(\,\vect{k} \cdot \,\vect{q})^2}{k^2} \Bigr)^2
   \nonumber \\ && \hspace{2cm} \times
    {\rm Re} \Bigl[~G^+_{~q}(\tau_1,~\tau_2)
    G^+_{~|\,\vect{k} - \,\vect{q}|}(\tau_1,~\tau_2)~\Bigr]
 \,.
  \nonumber \\ \label{Jsp}
\end{eqnarray}

Note that to compute the correlation function for the tensor
perturbation, $H_{T\,\vect{k}}^{(t)} (\tau)$ at conformal time $\tau$, 
it is necessary to consider the correlation function of
$p \pi_{\xi\,\vect{k}}$ for two different times $\tau_1$ and
$\tau_2$. This is because, as seen from Eq. (\ref{equation for HT2}),
 the expression
of $H_T^{(t)}$ includes the time integral. Therefore, there are two different
comoving horizon scales corresponding to the different times $\tau_1$ and $\tau_2$. 
For the same reason as in the case of
the scalar perturbations, we have introduced the IR
cut-off $H_i$. Nevertheless, we can see later, for the tensor
perturbations, even if we remove
the IR cut-off, the loop corrections remains finite.

By virtue of the same argument as that presented in the scalar
perturbations, $ J_{\rm sb}(\tau_1, \tau_2,\,\vect{k})$ contains
only the decaying modes. To show this, 
note that if either $- k \tau_1$
or $- k \tau_2$ is larger than unity, 
it does not produce cumulative contributions
because of  the oscillation of the Hankel
function in subhorizon region, 
as mentioned in Eq. (\ref{equation for HT}).
Hence, it is sufficient to consider only the case where
both
$- k \tau_1$ and $- k \tau_2$ are smaller than unity. If
 $\tau_1 \geq \tau_2$, then the inner momentum $q$ is
larger than $\mathcal{H}_2 \simeq - 1/\tau_2$. Hence, this implies that
$q$ is larger than $k$.
Approximating $|\,\vect{k} - \,\vect{q}|$ as 
$q$, we find that 
$J_{\rm sb}(\tau_1, \tau_2,\,\vect{k})$ contains only the
decaying mode as we have shown in the scalar perturbations. 
 The same discussion is valid also in case of 
$\tau_2 \geq \tau_1$. Hence, in order to compute the correlation
function of $p \pi_{\xi}^{(t)}$, it is sufficient to consider
only the contribution from the superhorizon region, 
$J_{\rm sp}(\tau_1, \tau_2,\,\vect{k})$.

As with the case of scalar perturbations, substituting the approximation of
the Wightman function in the superhorizon region into the definition of
$J_{\rm sp}(\tau_1, \tau_2,\,\vect{k})$, 
we find the contribution from the superhorizon
region  as
\begin{eqnarray}
 && J_{\rm sp}(\tau_1, \tau_2,\,\vect{k}) \equiv
 \frac{1}{4}
 \frac{~(\tau_1 \tau_2)^{1 - 2 \beta}}{(a_1 a_2)^2}
 \nonumber \\ &&~~~~
 \cdot 
 \Bigl[~ \theta(\tau_1 - \tau_2 )
   \int_{q \in [H_i, \mathcal{H}_2]} d^3 \,\vect{q}
   ~\Bigl( q^2 - \frac{(\,\vect{k} \cdot \,\vect{q})^2}{k^2} \Bigr)^2
 \nonumber \\ &&  \hspace{2cm} \times  
  \frac{1}{q^{3 + 2 (\varepsilon - \eta_V)~}
 |\,\vect{q} - \,\vect{k}|^{3 + 2 (\varepsilon - \eta_V)}}
 \nonumber \\[0.1cm] &&~~~~~~~
  + \theta(\tau_2 - \tau_1 )
   \int_{q \in [H_i, \mathcal{H}_1]} d^3 \,\vect{q}
   ~\Bigl( q^2 - \frac{(\,\vect{k} \cdot \,\vect{q})^2}{k^2} \Bigr)^2
 \nonumber \\ &&  \hspace{2cm} \times  
  \frac{1}{q^{3 + 2 (\varepsilon - \eta_V)~}
 |\,\vect{q} - \,\vect{k}|^{3 + 2 (\varepsilon - \eta_V)}} \Bigr]
\,.
 \nonumber \\ 
\end{eqnarray}
This loop integral is given by Eq. (\ref{solution for g}) 
in Appendix C.
It implies 
\begin{eqnarray}
 && J_{\rm sp}(\tau_1,~\tau_2,\,\vect{k}) \nonumber \\ 
 &&~ \simeq
  \frac{8 \pi}{15}  ~H_k^4 k 
 \Bigl[
  \theta(x_2 - x_1)~x_1^{2 \eta_V} x_2^{ -1 + 4 \varepsilon - 2 \eta_V} 
 \nonumber \\&& \hspace{1cm}  
 + \theta(x_1 - x_2)~x_1^{ -1 + 4 \varepsilon - 2 \eta_V} x_2^{2 \eta_V}  
 - \frac{3}{4} (x_1 x_2)^{2 \eta_V} \Bigr]
\,,
 \nonumber \\
\end{eqnarray}
where we have used
$H(\tau)^2 x^{- 2 \varepsilon}= H_k^2$. To derive this relation, we 
have taken 
the limit of $H_i \rightarrow 0$. No divergence appears. It
means that the cut-off for the infrared region is not necessary.
 It is interesting to note that  scalar
perturbations have the infrared divergence, but
tensor perturbations do not
suffer from the infrared problem. In Sec.\ref{Discussion}, we
discuss the origin of this infrared divergence and the reason why only
tensor perturbations does not contain it.

As a result, we obtain
 the correlation function for the tensor part of the
anisotropic pressure as  
\begin{eqnarray}
 && \langle p \pi_{\xi \,\vect{k}}(\tau_1) e^i_{~j} (\,\vect{k})
  ~p \pi_{\xi \,\vect{p}}(\tau_2) e^{j}_{~i} (\,\vect{p}) \rangle^{(4)}
 \nonumber \\ &&~~
 \simeq \frac{1}{4}~ (a_1 a_2)^{-2}~\delta (\,\vect{k} + \,\vect{p})
 ~J_{\rm sp}(\tau_1, \tau_2,\,\vect{k})
 \nonumber \\ &&~~
 \simeq   \frac{2 \pi}{15}
 ~\frac{H_k^4 k}{(a_1 a_2)^2} ~\delta (\,\vect{k} + \,\vect{p})
  \Bigl[
  \theta(x_2 - x_1) x_1^{2 \eta_V} x_2^{ -1 + 4 \varepsilon - 2 \eta_V}  
 \nonumber \\ && \hspace{1cm} 
 + \theta(x_1 - x_2) x_1^{ -1 + 4 \varepsilon - 2 \eta_V} x_2^{2 \eta_V}  
 - \frac{3}{4} (x_1 x_2)^{2 \eta_V} \Bigr]
\,.  \nonumber \\
 \label{correlation of ppif}
\end{eqnarray}
In our computation, we have neglected the loop corrections from the
tensor perturbations, because the amplitude of power spectrum for the
tensor perturbations is smaller than that for the scalar perturbations
by order of the slow-roll parameter $\varepsilon$.

\section{Loop corrections to the Correlation functions}
\label{Loop corrections}
As shown in Eqs. (\ref{rel bt df rho}) and (\ref{equation for HT2}),
the leading parts of the correlation function of the density
perturbation in a flat-slicing $\delta_f$ 
and of the tensor perturbation $H_T^{(t)}$ 
are determined by the stochastic variables
$\delta \rho_{\xi}$
and $p \pi_{\xi}^{(t)}$. 
We then have calculated the
correlation functions of $\delta \rho_{\xi}$ and $p \pi_{\xi}$ for the
noise kernel. Combining these results, we 
shall evaluate the correlation functions of
$\delta_f$ and $H_T^{(t)}$.    

\subsection{Scalar perturbations}
To evaluate the correlation function of the
density perturbation in flat-slicing $\delta_f$,
focusing on the proper nonlinear effects, we neglect the contribution
from the product of linear perturbations. This density
perturbation is related to the curvature perturbation in uniform slicing
$\zeta$ by Eq. (\ref{relation bt zdf}). 
Hence, once we find the loop corrections
to $\delta_f$, we also obtain the loop corrections to
$\zeta$. The curvature perturbation $\zeta$ is proportional to the
gravitational potential in the late time of the universe
and it is directly related to
the fluctuation of the temperature of CMB. That is why it is important
for us 
to consider this gauge-invariant variable among scalar perturbations.

From Eq. (\ref{rel bt df rho}) and
(\ref{correlation of drhof}), the loop corrections to the correlation
function of the density perturbation are given by
\begin{eqnarray}
  && \langle \delta_{f \,\vect{k}}(\tau)
  \delta_{f \,\vect{p}}(\tau) \rangle^{(4)}
  \nonumber \\ && 
  ~\simeq \frac{\pi}{2}
 \frac{(\kappa H_k)^4}{k^3} \eta_V^2
 (- k \tau)^{4 \eta_V}~
 \delta (\,\vect{k} + \,\vect{p})
 \nonumber \\  && \hspace{1cm}\times
 ~\Bigl[
 \frac{1}{3} - 
 \frac{1- (k/H_i)^{2(\varepsilon - \eta_V)}}{2(\varepsilon
 - \eta_V)}  \Bigr]
\,,
\label{loop correction with decay}
\end{eqnarray}
where we have used the relation of
$H(\tau)^2 x^{- 2 \varepsilon}= H_k^2$. In our previous work
\cite{YU20071}, we showed that when we solve explicitly the Einstein-
Langevin equation \cite{Martin:1999ih}, which includes an iterative aspect, it disturbs the
constant evolution of $\zeta$ in superhorizon region. Since this is a
problem of the way to quantize the scalar field and the gravitational field, the same effects may exist in the
present loop corrections. Hence, we restrict our discussions to the case
when $\eta_V \log x$ is smaller than unity, i.e., we assume that 
$(- k \tau)^{4 \eta_V}\approx 1$. Then, we find the
correlation function of the density perturbation $\delta_f$ as
\begin{eqnarray}
 && \langle \delta_{f  \,\vect{k}}(\tau)
  \delta_{f  \,\vect{p}}(\tau) \rangle^{(4)}
 \nonumber \\ &&
 \simeq \frac{\pi}{2}
 \frac{(\kappa H_k)^4}{k^3} \eta_V^2
  \delta (\,\vect{k} + \,\vect{p})~
 \Bigl\{
 \frac{1}{3} - \frac{1- (k/H_i)^{2(\varepsilon - \eta_V)}}{2(\varepsilon
 - \eta_V)}  \Bigr\} \,.  \nonumber \\
\end{eqnarray}
Taking into account Eq. (\ref{relation bt zdf}),
we obtain  the loop corrections to the correlation
function of the curvature perturbation in uniform density slicing
$\zeta$ as
\begin{eqnarray}
 && \langle \zeta_{ \,\vect{k}}(\tau)
  \zeta_{ \,\vect{p}}(\tau) \rangle^{(4)}
 \nonumber \\ &&
  \simeq \frac{\pi}{8}~
 \frac{(\kappa H_k)^4}{k^3}~ \Bigl( \frac{\eta_V}{\varepsilon} \Bigr)^2~
 \delta (\,\vect{k} + \,\vect{p})~
 \Bigl\{
 \frac{1}{3} - \frac{1- (k/H_i)^{2(\varepsilon - \eta_V)}}{2(\varepsilon
 - \eta_V)}  \Bigr\}
 \,. \nonumber \\ 
 \label{correlation of zeta1}
\end{eqnarray} 
The final result depends on the initial Hubble horizon scale,
$\mathcal{H}_i = H_i$, which is introduced to remove the infrared
divergence. The case with
$2|\varepsilon - \eta_V| \log(k/\mathcal{H}_i) < 1$ 
is particularly interesting.
This, in other words, corresponds to
the case of $N_k < 1/2 |\varepsilon - \eta_V|$,
where $N_k  \simeq \log(k/\mathcal{H}_i)$ is
 the e-folding from the beginning of inflation to the horizon
crossing time. In this case, this
correlation function is approximated as
\begin{eqnarray}
 && \langle \zeta_{\,\vect{k}}(\tau)
  \zeta_{\,\vect{p}}(\tau) \rangle^{(4)}
 \nonumber \\ &&~~
  \simeq \frac{\pi}{8}~
 \frac{(\kappa H_k)^4}{k^3}~ \Bigl( \frac{\eta_V}{\varepsilon} \Bigr)^2~
 \delta (\,\vect{k} + \,\vect{p})~ \Bigl(
 \frac{1}{3} + N_k  \Bigr) \,.
\end{eqnarray}
Note that there appears the logarithmic corrections. These results imply
that although the one-loop correction is suppressed by
$(\kappa H_k)^4$ and is smaller by the order of the $(\kappa H_k)^2$
than tree-level effects,
 it is amplified by the e-folding $N_k$ from the initial time
to the horizon crossing time, which can become large contrary to the
e-folding from the horizon crossing time to the end of the
inflation. However note that 
this amplification is derived by introducing the IR cut-off and the
obtained loop corrections significantly depend on the choice of the IR cut-off.

\subsection{Tensor perturbations}
The tensor perturbation
$H_T^{(t)}$ is related to the source term $p \pi_{\xi}^{(t)}$,
 and the correlation function of
$p \pi_{\xi}^{(t)}$ is given by Eq. (\ref{correlation of ppif}). Then,
integrating over $x=- k \tau$, we obtain the correlation function
and the amplitude of the tensor perturbation, which 
could be amplified by the
quantum effect of a scalar field. To integrate over $x$, it
is helpful to note the asymptotic behaviour of the Hankel function when
the argument $x$ is smaller than unity. As summarized in Appendix B in
\cite{YU20071}, the part of the integrand is approximated as
\begin{eqnarray}
  {\rm Im} \Bigl[ H_{\nu}^{~(1)}(x)~H_{\nu}^{~(2)}(x_1) \Bigr]
  \simeq - \frac{2}{3 \pi}
   \Bigl[\{ \Bigl( \frac{x_1}{x} \Bigr)^{\nu}  - 
  \Bigl( \frac{x}{x_1} \Bigr)^{\nu} \Bigr] \,.
\end{eqnarray} 
Using this approximation, we find 
the loop corrections to the correlation
function of the tensor perturbations  as
\begin{eqnarray}
 && \langle H_{T \,\vect{k}}^{(t)} (\tau) e^i_{~j}(\,\vect{k})~
 H_{T \,\vect{p}}^{(t)} (\tau) e^j_{~i}(\,\vect{p}) \rangle^{(4)}
 \nonumber \\ &&~~
 \simeq  \frac{\pi}{135} \frac{(\kappa H_k)^4}{k^3}
 ~\delta (\,\vect{k} + \,\vect{p})~
 \Bigl(~ \frac{7}{6} - 15~ x^{2 + 2 \eta_V} \Bigr) \,.
 \nonumber \\
 \label{correlation of HT1}
\end{eqnarray}
It is interesting to note that there is no dependence on the infrared
cut off $H_i$ in the tensor perturbations. Furthermore, the tensor
perturbations are divergence free in the superhorizon region.  
We shall discuss the reason in the next section.  
In our computation, we have neglected the loop corrections from the tensor perturbations, since the amplitude of the tensor perturbations are smaller than that of the scalar perturbations by order of the slow-roll parameter $\varepsilon$. 

\section{Discussions} \label{Discussion}

Using the Einstein-Langevin equation proposed in \cite{Martin:1999ih}, we
calculate the loop corrections to the scalar perturbations and the
 tensor perturbations, which
 are amplified through the nonlinear
 interaction between the scalar field and the gravitational field. 
Here we discuss the origin of the amplification.

When we consider the loop corrections in inflationary universe, there
are two different divergences. One is the ultraviolet (UV) divergence.
 Since
this divergence is originated by short wave modes, such divergence
also appears in the quantum field theory in a 
Minkowski background.
In inflationary spacetime, there exists another divergence, which 
is not found in Minkowski spacetime. This is the IR divergence. 
To avoid this IR divergence, we have
introduced the cut-off at the initial Hubble horizon size. Then the
amplitude of the one-loop corrections to the curvature perturbation
$\zeta$ is amplified by the e-folding from the initial time of inflation
to the horizon crossing time, i.e., the logarithmic correction. If this is
true, this amplification may make it possible to detect these loop
corrections.  Then it will be 
a great help to clarify the fundamental properties of
an inflation model.

So far we have several discussions about this
logarithmic corrections due to IR divergence.
Early works about this problem are done by Boyanovsky, de Vega
and Sanchez 
\cite{Boyanovsky:2004gq, Boyanovsky:2004ph, Boyanovsky:2005sh,
Boyanovsky:2005px}.
They calculated one loop corrections by light scalar and fermion fields
to the inflaton potential, and also
evaluated those by the gauge invariant curvature and
tensor perturbations. 
They found that there appear the IR enhancements 
both in the  scalar field corrections 
and curvature perturbations, while
both fermion corrections and tensor perturbations
 do not exhibit IR divergences.
Weinberg also pointed that the loop corrections
to the primordial perturbations behave at most logarithmic
\cite{Weinberg:2005vy, Weinberg:2006ac}. Afterward  Sloth 
considered the loop corrections to the fluctuation of the scalar
field in flat-slicing \cite{Sloth:2006az, Sloth:2006nu}. To avoid
IR divergence, he introduced the cut off by the initial horizon
scale. As a result, he found that
 the loop correction is amplified by the e-folding from the
initial time of inflation to the horizon crossing time,
which is also found in this paper from the analysis based on stochastic
gravity. Following Sloth, Seery readdressed this problem more
carefully \cite{Seery:2007we, Seery:2007wf}. In particular, he analysed
the evolution in the superhorizon region using the $\delta N$  formula
\cite{Sasaki:1995aw, Wands:2000dp},
and improved his results.
In this paper, we have computed the loop corrections by stochastic
gravity, and found the similar logarithmic corrections for scalar
perturbations. 
The same logarithmic behaviours have been found in other interacting
systems \cite{Onemli:2002hr, Brunier:2004sb, Prokopec:2007ak}.
 However,
we should note that the IR problem requires the careful treatment and
the way to evaluate IR effects is controversial
 \cite{Garriga:2007zk, Tsamis:2007is}.

Recently, Lyth has claimed that, to avoid the 
assumptions on unknown parts of the
universe, the calculation about loop corrections should be done
inside a comoving box, whose size $L$ is not too much bigger
than the present horizon scale \cite{Lyth:2007jh}. 
The IR corrections are significantly reduced, although we 
still find the logarithmic behaviours.
Furthermore, Bartolo et al. claimed that a stochastic approach 
 plays a crucial role to deal with this problem \cite{Bartolo:2007ti}.
 In relation to their
insists, we should stress our interesting results.
That is, although the scalar
perturbations are amplified by the logarithmic corrections, the tensor
perturbations are not. Even if we remove the IR cut-off, the IR
divergence does not appear in tensor perturbations. This difference
between the scalar and tensor perturbations seems to
be related to the origin of these logarithmic corrections.

To consider the origin of this logarithmic corrections due to
the IR cut-off, we first
 mention the prediction in stochastic
inflation\cite{Starobinsky:1986fx, Nakao:1988yi, Nambu:1988je, Morikawa:1987ci,
Tanaka:1997iy, Matarrese:2003ye, Liguori:2004fa, Rigopoulos:2004gr, Rigopoulos:2004ba, Rigopoulos:2005xx,
Rigopoulos:2005ae}. 
 In stochastic inflation,
 the long wave mode $k$ with $k < aH$ of the scalar field couples to
the short wave mode $k$ with $k > aH$ through the nonlinear 
self-interaction of the scalar field.
Then the long wave modes
are affected by the quantum fluctuation of the short wave modes. 
As a result, the long wave modes come
to show stochastic behavior.
This stochastic behavior of the long wave modes affects 
the background quantities.

In our case, due to the nonlinear interaction between the
gravitational field and the scalar field, the long wave modes and the
background quantities come to show the stochastic behavior. Since the scalar perturbations are defined
as the deviation from the background quantities, the stochastic
fluctuation of the background quantities affect the behavior of the
perturbed variables. As a result, it induces the logarithmic secular
evolution of the perturbed variables. On the other hand, there are no
background tensor modes, and then the one loop corrections to
the tensor perturbation can avoid being affected by the background
stochastic fluctuations. Furthermore, although in this paper, as the simplest step for 
treating the IR divergence, we have simply neglected the long wave modes with
 $- k \mathcal{H}_i > 1$, the infrared modes require the more careful
 treatment. We will propose the way to regularize the IR corrections in \cite{YU20081}.

There is another notable difference between the scalar and the tensor
 perturbations. As pointed out in our previous work, in the present
 approach of stochastic gravity the longitudinal part of gravitational
 field is included iteratively. This affects the behavior of
 perturbations in the superhorizon region. In particular, the curvature
 perturbation deviates from constant when the e-folding from the horizon
 crossing time exceeds the definite critical value
(~$|\mbox{slow-roll parameter}|^{-1}$). Since this is the problem of the
 way to quantize the gravitational field and the matter field, the loop corrections to the scalar
perturbations are also influenced by the nonexistence of the
longitudinal part of the quantized gravitational field. In fact, as
shown in Eq. (\ref{loop correction with decay}), the one loop correction
to the curvature perturbation $\zeta$ evolves as $x^{4 \eta_V}$ in the
superhorizon region. On the other hand, as shown in Eq. 
(\ref{correlation of HT1}), the tensor
perturbations do not decay. This means that even if we use the
 Einstein-Langevin equation in the present
 iterative way, it does not affect the one loop correction to the tensor
 perturbations.

\acknowledgments
We would like to thank B.L. Hu, A. Roura,
 M. Sasaki, J. Soda, A.A. Starobinsky, T. Tanaka, and  
 E. Verdaguer for valuable discussions. 
Y.U. would like to specially acknowledge B.L. Hu, A.A. Starobinsky, 
T. Tanaka, and E. Verdaguer for very fruitful
comments and suggestions.
This work was partially supported 
by the Japan Society for 
Promotion of Science (JSPS) Research Fellowships (Y.U.),
and by the Grant-in-Aid for Scientific Research
Fund of the JSPS (No.19540308) and for the
Japan-U.K. Research Cooperative Program,
and by the Waseda University Grants for Special Research Projects and
 for the 21st-Century
COE Program (Holistic Research and Education Center for Physics
Self-Organization Systems) at Waseda University.
We would also like to acknowledge the hospitality of the Gravitation and
 Cosmology Group at Barcelona University, where the early stage of the
 present work was done.

\appendix
\begin{widetext}
\section{Computations of the Noise kernel}  \label{computation of N}
In Appendix \ref{computation of N}, we calculate the noise kernel, which
is defined by Eq. (\ref{def of Noise kernel}) and 
(\ref{def of F}). As shown in Sec.\ref{Loop corrections}, we have to
compute the correlation function of the density perturbation and the
transverse traceless part of the anisotropic pressure of the stochastic variable
$\xi_{ab}$. Then, we compute the $(0,0,0',0')$ component  and the
transverse traceless part of the $(i,j,k',l')$ component of the noise
kernel, i.e.,  $F^{0~0'}_{~0~~0'}(x_1,x_2)$ and 
$F^{i~k'}_{~j~~l'}(x_1,x_2)$. Note that the noise kernel is computed
from the quantum fluctuation of the scalar field on the background
spacetime. Decomposing the scalar field into $\phi = \phi_{cl} + \psi$,
the energy momentum tensor 
\begin{eqnarray}
 && T_{ab} = \nabla_a \phi \nabla_b \phi - \frac{1}{2} g_{ab}
 [\nabla_c \phi \nabla^c \phi + 2 V(\phi)]
\label{Tab for single phi} 
\end{eqnarray}
is expressed as the classical part and the fluctuation part as follows:
\begin{eqnarray}
  T_{ab} &=& T_{ab}^{~(cl)} + \delta T_{ab}~;
 \nonumber \\
 && T_{ab}^{~(cl)} \equiv \delta^0_{~a} \delta^0_{~b} a^2 \dot{\phi}_{cl}^2 
  + \frac{1}{2} g_{ab} \dot{\phi}_{cl}^2 - g_{ab}  V(\phi_{cl}) \\
 && \delta T_{ab} \equiv (\delta^0_{~a} \nabla_b \psi + \delta^0_{~b}
  \nabla_a \psi + \eta_{ab} \nabla_0 \psi ) a \dot{\phi}_{cl}
 + \nabla_a \psi \nabla_b \psi
 - \frac{1}{2} g_{ab} \nabla_c \psi \nabla^c \psi
- g_{ab}  V(\phi_{cl}) \displaystyle \sum_{m=1}^{\infty}
  \frac{\alpha^{(m)}}{m!}  (\kappa \psi)^m 
\,.
\label{def of deltaT}
\end{eqnarray}
The noise kernel, which represents the fluctuation of the
energy-momentum tensor, can be expressed in terms of $\delta T_{ab}$. In
fact, substituting this decomposed energy-momentum tensor into the
definition of $F_{abc'd'}(x_1, x_2)$, we can express the two-point function
$F_{abc'd'}(x_1, x_2)$ as
\begin{eqnarray}
&&\hspace{-0.75cm} F_{abc'd'}(x_1, x_2) 
\nonumber 
\\
&=& \langle \widehat{\delta T}_{ab}(x_1)
 \widehat{\delta T}_{c'd'}(x_2) \rangle
 - \langle \widehat{\delta T}_{ab} (x_1) \rangle 
 \langle \widehat{\delta T}_{c'd'} (x_2) \rangle
 \nonumber \\[0.1cm] 
 &=& a_1 a_2 \dot{\phi}_{cl,1}  \dot{\phi}_{cl,2}
 ~[~\delta^0_{~a} \delta^0_{~c'} G^{H}_{;bd'}
 + \delta^0_{~a} \delta^0_{d'} G^{H}_{;bc'}
 + \delta^0_{~b} \delta^0_{c'}G^H_{;ad'}
 + \delta^0_{~b} \delta^0_{d'}G^H_{;ac'} \nonumber \\[0.1cm]
 && \hspace{3cm} +~ \eta_{c'd'}( \delta^0_{~a} G^H_{;b0'}
 + \delta^0_{~b} G^H_{;a0'} )
 + \eta_{ab}( \delta^{0'}_{~c'} G^H_{;0d'}
 + \delta^{0'}_{~d'} G^H_{;0c'})
 + \eta_{ab} \eta_{c'd'} G^H_{;00'}]
 \nonumber \\[0.1cm] && \hspace{0.25cm}
 +~ G^H_{;ac'} G^H_{;bd'}
 + G^H_{;ad'} G^H_{;bc'}
 - (a_1)^2 \eta_{ab} G^H_{;ec'} G^{H;e}_{;d'}
 - (a_2)^2 \eta_{c'd'} G^H_{;af'} G^{H;f' }_{;b}
 + \frac{(a_1 a_2)^2}{2} \eta_{ab} \eta_{c'd'} G^H_{;ef'} G^{H;ef'}
\nonumber \\[0.1cm] && \hspace{0.25cm}
 -~ (a_1)^2 \eta_{ab} V_{cl,1} \Bigl[
 a_2 \dot{\phi}_{cl,2}~ \tilde{\alpha}_1^{(1)} \kappa
 (\delta^{0'}_{~c'} G^H_{;d'} + \delta^{0'}_{~d'} G^H_{;c'}
 + \eta_{c'd'} G^H_{;0'})
 + \tilde{\alpha}^{(2)}_{1} \kappa^2 \Bigl( G^H_{;c'} G^H_{;d'}
 - \frac{(a_2)^2}{2}~ \eta_{c'd'} G^H_{;f'} G^{H;f'} 
 \Bigr) \Bigr]
\nonumber \\[0.2cm] && \hspace{0.25cm}
 -~(a_2)^2 \eta_{c'd'} V_{cl,2} \Bigl[
 a_1 \dot{\phi}_{cl,1} ~\tilde{\alpha}_2^{(1)} \kappa
 (\delta^0_{~a} G^H_{;b} + \delta^0_{~b} G^H_{;a}
 + \eta_{ab} G^H_{;0})
 + \tilde{\alpha}^{(2)}_{2}
  \kappa^2 \Bigl( G^H_{;a} G^H_{;b}
  - \frac{(a_1)^2}{2}~\eta_{ab} G^H_{;e} G^{H;e} \Bigr) \Bigr]
    \nonumber \\[0.1cm] && \hspace{0.25cm}
   +~(a_1 a_2)^2 \eta_{ab} \eta_{c'd'} V_{cl,1} V_{cl,2}
 \Bigl[
 \tilde{\alpha}^{(1)}_1 \tilde{\alpha}^{(1)}_2 \kappa^2 G^H 
 + \frac{\tilde{\alpha}^{(2)}_1 \tilde{\alpha}^{(2)}_{2}}{2!}(\kappa^2 G^H)^2 
 + \frac{\tilde{\alpha}^{(3)}_{1} \tilde{\alpha}^{(3)}_{2}}{3!}(\kappa^2 G^H)^3
 + O \Bigl((\varepsilon_{SR}^{1/2} \kappa H )^{8} \Bigr) \Bigr]
\,,
  \nonumber \\ \label{Fabcd} 
\end{eqnarray}
where
$G^H \equiv \langle \Omega| \hat{\psi}_H(x_1) \hat{\psi}_H(x_2) | 
\Omega \rangle$
is the Wightman Green function for the interacting system, 
which is defined as the two-point function of the Heisenberg field $\hat{\psi}_H$.
Here we have redefined the coefficients $\tilde{\alpha}^{m}$, including the
divergent part $G^{H}(x,x)$ as follows:
\begin{eqnarray}
 && \tilde{\alpha}^{(1)}_x \equiv 
 \alpha^{(1)}_x + \frac{ \alpha^{(3)}_x}{2} \kappa^2 G^H_{xx}
 + \frac{ \alpha^{(5)}_{x}}{8} \Bigl\{\kappa^2 G^H_{xx} \Bigr\}^2
 + O\Bigl((\varepsilon_{SR}^{1/2} \kappa H )^6 \Bigr)
 \nonumber \\
 && \tilde{\alpha}^{(2)}_x \equiv
  \alpha^{(2)}_x + \frac{\alpha^{(4)}_x}{2} \kappa^2 G^H_{xx} +
  O(\varepsilon_{SR}^3 (\kappa H)^4) \nonumber \\
 && \tilde{\alpha}^{(3)}_x = \alpha^{(3)}_x + O \Bigl((\varepsilon_{SR}^{1/2}
  \kappa H )^2 \Bigr)  \,.
\end{eqnarray}
Strictly speaking, we have to renormalize
these divergent terms into the coefficients of the potential
$V(\phi)$. In this paper, we assume that these divergent parts are
removed by an appropriate
 renormalization procedure. So the finite part of these
radiative correction terms is much smaller than the leading term among
the coefficients $\tilde{\alpha}^{(m)}$. Here we neglect them and
approximate $\tilde{\alpha}^{(m)}$ as $\alpha^{(m)}$.

In our previous work \cite{YU20071}, we discussed the linear
perturbations which are proportional to $(\kappa H)^2$. In this paper,
we consider the leading loop corrections which are proportional to
$(\kappa H)^4$. The self-interaction part contributes to the
effective action from the order $(\kappa H)^6$, which is depicted in 
Fig. \ref{fig.eps} (3). In these diagrams, the solid line represents the
interacting picture field which satisfies the equation
\begin{eqnarray}
 [\partial^2_{~0} + (D-2) \mathcal{H} \partial_0
 - \nabla^2 + a^2 V''(\phi_{cl})]~\psi_f (x) = 0  \,.
\end{eqnarray}
When we compute the effective action up to the order of $(\kappa H)^4$, we
can replace the Wightman function for the Heisenberg field
$G^H (x_1, x_2)$ to the Wightman function for the interaction picture field
$G^+ (x_1, x_2)$. Then, the parts of $F_{abc'd'}(x_1, x_2)$, whose orders are
$(\kappa H)^2$ and $(\kappa H)^4$, are given by
\begin{eqnarray}
 F^{(2)}_{abc'd'}(x_1, x_2)  &=& 
 a_1 a_2 \dot{\phi}_{cl,1}  \dot{\phi}_{cl,2}
 ~[~\delta^0_{~a} \delta^0_{~c'} G^+_{;bd'}
 + \delta^0_{~a} \delta^0_{d'} G^+_{;bc'}
 + \delta^0_{~b} \delta^0_{c'} G^+_{;ad'}
 + \delta^0_{~b} \delta^0_{d'} G^+_{;ac'} \nonumber \\[0.1cm]
 && \hspace{3cm} +~ \eta_{c'd'}( \delta^0_{~a} G^+_{;b0'}
 + \delta^0_{~b} G^+_{;a0'} )
 + \eta_{ab}( \delta^{0'}_{~c'} G^+_{;0d'}
 + \delta^{0'}_{~d'} G^+_{;0c'})
 + \eta_{ab} \eta_{c'd'} G^+_{;00'}]  
 \nonumber \\[0.1cm] && \hspace{0.5cm}
 -~(a_1)^2 \eta_{ab} V_{cl,1}  a_2 \dot{\phi}_{cl,2}~\alpha^{(1)}_1
 ~ \kappa
 (\delta^{0'}_{~c'}~ G^+_{;d'} + \delta^{0'}_{~d'}~ G^+_{;c'}
 + \eta_{c'd'}~ G^+_{;0'}) \nonumber \\[0.1cm]
 && \hspace{0.5cm}
 -~(a_2)^2 \eta_{c'd'} V_{cl,2} a_1 \dot{\phi}_{cl,1} \alpha^{(1)}_{2}
 ~ \kappa (\delta^0_{~a}~ G^+_{;b} + \delta^0_{~b}~ G^+_{;a}
 + \eta_{ab}~ G^+_{;0}) \nonumber \\[0.1cm] && \hspace{0.5cm}
   +~(a_1 a_2)^2 \eta_{ab} \eta_{c'd'} V_{cl,1} V_{cl,2} ~ \alpha^{(1)}_1
 \alpha^{(1)}_2 \kappa^2 G^+ \label{F2abcd}
\end{eqnarray}
and
\begin{eqnarray}
 F^{(4)}_{abc'd'}(x_1, x_2)  &=&
 G^+_{;ac'} G^+_{;bd'}
 + G^+_{;ad'} G^+_{;bc'}
 - (a_1)^2 \eta_{ab} G^+_{;ec'} G^{;e}_{~~;d'}
 - (a_2)^2 \eta_{c'd'} G^+_{;af'} G^{+~;f' }_{;b}
 + \frac{(a_1 a_2)^2}{2} \eta_{ab} \eta_{c'd'} G^+_{;ef'} G^{+;ef'}
\nonumber \\[0.1cm] && \hspace{0.5cm}
 -~(a_1)^2 \eta_{ab} V_{cl,1}  \alpha^{(2)}_1
  \kappa^2~\Bigl\{~ G^+_{;c'} G^+_{;d'}
 - \frac{1}{2}~(a_2)^2 \eta_{c'd'} G^+_{;f'} G^{+;f'} ~\Bigr\}
\nonumber \\ && \hspace{0.5cm}
 -~(a_2)^2 \eta_{c'd'} V_{cl,2} \alpha^{(2)}_2
 \kappa^2~\Bigl\{~ G^+_{;a} G^+_{;b}
  - \frac{1}{2}~(a_1)^2 \eta_{ab} G^+_{;e} G^{+;e}~\Bigr\}
 \nonumber \\[0.1cm] && \hspace{0.5cm}
   +~ (a_1 a_2)^2 \eta_{ab} \eta_{c'd'} V_{cl,1} V_{cl,2}~
 \frac{\alpha^{(2)}_1 \alpha^{(2)}_2}{2}~
 (\kappa^2 G^+)^2
\,,  \label{F4abcd}
\end{eqnarray}
respectively,
where the superscript numbers represent the powers of $(\kappa H)$.
To compute the correlation functions of the primordial
perturbations in momentum space, it is convenient to use the 
Fourier-transformed Wightman function given by
\begin{eqnarray}
 G_k^+ (\tau_1,~\tau_2) \equiv
 \psi_{f, \,\vect{k}}(\tau_1) \psi^*_{f, \,\vect{k}}(\tau_2) 
\,,
\end{eqnarray}
where the mode function $\psi_{f, \,\vect{k}} (\tau)$ satisfies
\begin{eqnarray}
 [\partial^2_{~0} + 2 \mathcal{H} \partial_0 +
 k^2 + a^2 V(\phi_{cl}) \eta_V \kappa^2 ]~\psi_{f, \,\vect{k}} (\tau) = 0
\,.
  \label{eq for psifk}
\end{eqnarray}

\subsection{Scalar perturbations}
Using Eq. (\ref{F4abcd}), we give the leading loop
corrections to the correlation function of the stochastic variable
$\xi_{ab}$. To compute the loop corrections to the scalar
perturbations, first we have to find the correlation function of
the density perturbation of the stochastic variable, which is given by
the $(0,0,0',0')$ component of the noise kernel. It is obtained
from the real part of $F_{000'0'}$. The part of order
of $(\kappa H)^4$ is
\begin{eqnarray}
  \hat{F}^{(4)0~~0'}_{~~~~~~0~~0'} (\tau_1,~\tau_2,\,\vect{k},\,\vect{p})
  &\equiv& \int d^3 \,\vect{x}_1  \int d^3 \,\vect{x}_2
  e^{-i \,\vect{k} \cdot \,\vect{x}_1} e^{-i \,\vect{p} \cdot \,\vect{x}_2}
  F^{(4)0~~0'}_{~~~~~~0~~0'}(x_1,~x_2)  \nonumber \\
  &=& \frac{1}{2}~\delta^{(3)} (\,\vect{k} + \,\vect{p})
  \int d^3 \,\vect{q} \Bigl\{ (a_1)^{-2}
  \Bigl( \partial^{~~q}_{\tau_1} \partial^{~~k-q}_{\tau_1}
  - \,\vect{q} \cdot (\,\vect{k} - \,\vect{q}) \Bigr)
  + \alpha_1^{(2)} V_{cl,1} \kappa^2 \Bigr\}
  \nonumber \\ && \hspace{1cm} \times
  \Bigl\{ (a_2)^{-2}
  \Bigl( \partial^{~~q}_{\tau_2} \partial^{~~k-q}_{\tau_2}
  - \,\vect{q} \cdot (\,\vect{k} - \,\vect{q}) \Bigr)
  + \alpha_2^{(2)} V_{cl,2} \kappa^2 \Bigr\}
  ~G^+_{~q}(\tau_1,~\tau_2) G^+_{~|\,\vect{k} - \,\vect{q}|}(\tau_1,~\tau_2)
 \nonumber   \label{F0000 in momentum sp} \\
\end{eqnarray}
We put the momentum superscript on the partial derivative operator to represent
that $\partial^{~~q}_{\tau_1}$ operates only to the Wightman
function, $G^+_{~q}(\tau_1,~\tau_2)$.
For example, 
$\partial^{~~q}_{\tau_1} \partial^{~~k-q}_{\tau_1} ~G^+_{~q}(\tau_1,~\tau_2)
G^+_{~|\,\vect{k} - \,\vect{q}|}(\tau_1,~\tau_2)$ means $ \partial_{\tau_1}
~G^+_{~q}(\tau_1,~\tau_2) \partial_{\tau_1} G^+_{~|\,\vect{k} - \,\vect{q}|}
(\tau_1,~\tau_2)$. Taking into account that the correlation function of
the stochastic variable is given by the noise kernel
(\ref{def of Noise kernel}), we find Eq. (\ref{F0000 in momentum sp}) implies
\begin{eqnarray}
  \langle \delta \rho_{\xi \,\vect{k}}(\tau_1)
   \delta \rho_{\xi \,\vect{p}}(\tau_2)  \rangle^{(4)} 
  &=& \int d^3 \,\vect{x}_1  \int d^3 \,\vect{x}_2
  e^{-i \,\vect{k} \cdot \,\vect{x}_1} e^{-i \,\vect{p} \cdot \,\vect{x}_2}
 \langle \xi^0_{~0}(x_1) \xi^{0'}_{~~0'}(x_2) \rangle^{(4)}
 \nonumber \\
 &=& \frac{1}{8}  \Bigl[
 \hat{F}^{(4)0~~0'}_{~~~~~~0~~0'} (\tau_1,~\tau_2,\,\vect{k},\,\vect{p})
 +  \hat{F}^{(4)0~~0'}_{~~~~~~0~~0'}
 (\tau_1,~\tau_2,~-\,\vect{k},~-\,\vect{p})^{*} \Bigr] \nonumber \\
 &=& \frac{1}{8}~\delta^{(3)} (\,\vect{k} + \,\vect{p})
 \int d^3 \,\vect{q} \Bigl\{ (a_1)^{-2}
 \Bigl( \partial^{~~q}_{\tau_1} \partial^{~~k-q}_{\tau_1}
 - \,\vect{q} \cdot (\,\vect{k} - \,\vect{q}) \Bigr)
 + \alpha_1^{(2)} V_{cl,1} \kappa^2 \Bigr\}
 \nonumber \\ && \hspace{1cm} \times
 \Bigl\{ (a_2)^{-2}
 \Bigl( \partial^{~~q}_{\tau_2} \partial^{~~k-q}_{\tau_2}
 - \,\vect{q} \cdot (\,\vect{k} - \,\vect{q}) \Bigr)
 + \alpha_2^{(2)} V_{cl,2} \kappa^2 \Bigr\}
 {\rm Re} \Bigl[ ~G^+_{~q}(\tau_1,~\tau_2)
 G^+_{~|\,\vect{k} - \,\vect{q}|}(\tau_1,~\tau_2) \Bigr]  \,.
 \nonumber \\   \label{correlation of rho}
\end{eqnarray}
On the third equality, we have transformed the momentum 
$\,\vect{q}$ to $-
\,\vect{q}$ in the integral of $\hat{F}^{(4)0~~0'}_{~~~~~~0~~0'}
(\tau_1,~\tau_2,~-\,\vect{k},~-\,\vect{p})^{*}$. This integral corresponds to
the integral of the momentum of the internal line of the loop graph.

\subsection{Tensor perturbations}
Next we calculate the correlation function of the transverse
traceless part of the anisotropic pressure for which 
we have to compute
the loop corrections to the tensor perturbations. The correlation
function of the pressure part of the stochastic variable is given by the
bi-tensor $F^{(4)i~~l'}_{~~~~~j~~m'}$ in momentum space as
\begin{eqnarray}
 \hat{F}^{(4)i~~l'}_{~~~~~j~~m'} (\tau_1,~\tau_2,\,\vect{k},\,\vect{p})
 &=& \int d^3 \,\vect{x}_1  \int d^3 \,\vect{x}_2 
  e^{-i \,\vect{k} \cdot \,\vect{x}_1} e^{-i \,\vect{p} \cdot \,\vect{x}_2}
  F^{(4)i~~l'}_{~~~~~j~~m'} (x_1,~x_2)  \nonumber \\
 &=& \frac{1}{2}~ (a_1 a_2)^{-2}~\delta (\,\vect{k} + \,\vect{p})
 \int d^3 \,\vect{q}~[
 \{q^i (k_j - q_j) + q_j (k^i - q^i) \}
 \{ q^{l'}(k_{m'} - q_{m'}) + q_{m'} (k^{l'} - q^{l'}) \}
 \nonumber \\ && \hspace{5cm} +~ (\mbox{trace part})~]~
  G^+_{~q}(\tau_1,~\tau_2)
 G^+_{~|\,\vect{k} - \,\vect{q}|}(\tau_1,~\tau_2)
\,.
\end{eqnarray}
The transverse traceless part of any tensor on spatial flat
hypersurface can be extracted by means of the projection operator,
$P^{i}_{~j}(\,\vect{k})$, as follows, 
\begin{eqnarray}
 t^{~i}_{t~~j} (\,\vect{k})
 = \Bigl[ P^{i}_{~k}(\,\vect{k}) P^{l}_{~j}(\,\vect{k})
 - \frac{1}{2} P^{i}_{~j}(\,\vect{k})P^{l}_{~k}(\,\vect{k}) \Bigr]~
 t^k_{~l}(\,\vect{k}) 
\hspace{1.5cm} 
 P^{i}_{~j}(\,\vect{k}) \equiv \delta^i_{~j} - \frac{k^i k_j}{k^2} \,.
\end{eqnarray}
Operating the projection operator onto
$\hat{F}^{(4)i~~l'}_{~~~~~j~~m'} (\tau_1,~\tau_2,\,\vect{k},\,\vect{p})$, we
extract its tensor part as follows:
\begin{eqnarray}
 [\hat{F}^{(4)i~~l'}_{~~~~~j~~m'} (\tau_1,~\tau_2,\,\vect{k},
\,\vect{p})]_{TT}  
 &=& \frac{1}{2 (a_1 a_2)^{2}} ~\delta (\,\vect{k} + \,\vect{p})
 \int d^3 \,\vect{q}
 ~G^+_{~q}(\tau_1,~\tau_2)
 G^+_{~|\,\vect{k} - \,\vect{q}|}(\tau_1,~\tau_2)
 \nonumber \\ && \hspace{0.25cm} \times 
 \Bigl\{- 2 q^i q_j 
 + \delta^i_{~j}
  \Bigl(q^2 - \frac{(\,\vect{q} \cdot \,\vect{k})^2}{k^2} \Bigr)
 - \frac{k^i k_j}{k^2} 
   \Bigl(q^2 + \frac{(\,\vect{q} \cdot \,\vect{k})^2}{k^2} \Bigr)
 + 2 \frac{\,\vect{q} \cdot \,\vect{k}}{k^2} (k^i q_j + q^i k_j) \Bigr\}
 \nonumber \\ && \hspace{0.25cm} \times 
 \Bigl\{- 2 q^{l'} q_{m'} 
 + \delta^{l'}_{~~m'}
  \Bigl(q^2 - \frac{(\,\vect{q} \cdot \,\vect{k})^2}{k^2} \Bigr)
 - \frac{k^{l'} k_{m'}}{k^2} 
   \Bigl(q^2 + \frac{(\,\vect{q} \cdot \,\vect{k})^2}{k^2} \Bigr)
 + 2 \frac{\,\vect{q} \cdot \,\vect{k}}{k^2}
  (k^{l'} q_{m'} + q^{l'} k_{m'}) \Bigr\} \,.
\nonumber \\
\end{eqnarray}
Consequently, taking into account the definition of the noise kernel
(\ref{def of Noise kernel}), we can give the correlation function of
the transverse traceless part of the anisotropic pressure
$p \tau^{(t)}_{~\xi}$ by
\begin{eqnarray}
 && \langle p \pi^{(t)}_{\xi \,\vect{k}}(\tau_1) e^i_{~j} (\,\vect{k})
  p \pi^{(t)}_{\xi\,\vect{p}}(\tau_2) e^{l'}_{~~m'} (\,\vect{p})
  \rangle^{(4)}
   \nonumber \\ && \hspace{1.5cm}
  = \int d^3 \,\vect{x}_1  \int d^3 \,\vect{x}_2
  e^{-i \,\vect{k} \cdot \,\vect{x}_1} e^{-i \,\vect{p} \cdot \,\vect{x}_2}
  ~ \langle \xi^i_{~j}(x_1) \xi^{l'}_{~~m'}(x_2) \rangle^{(4) tt}
  \nonumber \\ && \hspace{1.5cm}
  = \frac{1}{8}~ (a_1 a_2)^{-2}~\delta (\,\vect{k} + \,\vect{p})
  \int d^3 \,\vect{q}~
  Re \Bigl[~G^+_{~q}(\tau_1,~\tau_2)
  G^+_{~|\,\vect{k} - \,\vect{q}|}(\tau_1,~\tau_2)~\Bigr]
  \nonumber \\ && \hspace{2.5cm} \times 
 \Bigl\{- 2 q^i q_j 
 + \delta^i_{~j}
  \Bigl(q^2 - \frac{(\,\vect{q} \cdot \,\vect{k})^2}{k^2} \Bigr)
 - \frac{k^i k_j}{k^2} 
   \Bigl(q^2 + \frac{(\,\vect{q} \cdot \,\vect{k})^2}{k^2} \Bigr)
 + 2 \frac{\,\vect{q} \cdot \,\vect{k}}{k^2} (k^i q_j + q^i k_j) \Bigr\}
 \nonumber \\ && \hspace{2.5cm} \times 
 \Bigl\{- 2 q^{l'} q_{m'} 
 + \delta^{l'}_{~~m'}
  \Bigl(q^2 - \frac{(\,\vect{q} \cdot \,\vect{k})^2}{k^2} \Bigr)
 - \frac{k^{l'} k_{m'}}{k^2} 
   \Bigl(q^2 + \frac{(\,\vect{q} \cdot \,\vect{k})^2}{k^2} \Bigr)
 + 2 \frac{\,\vect{q} \cdot \,\vect{k}}{k^2}
  (k^{l'} q_{m'} + q^{l'} k_{m'}) \Bigr\}  \,. \nonumber \\
\end{eqnarray}
Especially when we contract the suffices
$(i,~m')$ and $(j,~l')$, this correlation
function is rewritten as
\begin{eqnarray}
 &&  \langle p \pi_{\xi \,\vect{k}}(\tau_1) e^i_{~j} (\,\vect{k})
  ~p \pi_{\xi \,\vect{p}}(\tau_2) e^{j}_{~i} (\,\vect{p}) \rangle^{(4)}
  \nonumber \\ && \hspace{1cm}
 = \frac{1}{4}~ (a_1 a_2)^{-2}~\delta (\,\vect{k} + \,\vect{p})
 \int d^3 \,\vect{q}~
 \Bigl( q^2 - \frac{(\,\vect{k} \cdot \,\vect{q})^2}{k^2} \Bigr)^2
 ~Re \Bigl[~G^+_{~q}(\tau_1,~\tau_2)
 G^+_{~|\,\vect{k} - \,\vect{q}|}(\tau_1,~\tau_2)~\Bigr] \,.
 \label{correlation of ppi}
\end{eqnarray}

\section{Loop integration} \label{loop integration}
To integrate over the inner momentum, $\,\vect{q}$, it is
convenient to consider the functions, $f(k, \delta, \mathcal{H})$ and
$g(k, \delta, \mathcal{H})$ which are defined as
\begin{eqnarray}
 f(k, \delta, \mathcal{H}) &\equiv& \int_{q \in [k_i, \mathcal{H}]} d^3 q \frac{1}{q^{3
  + \delta}} \frac{1}{|\,\vect{k} - \,\vect{q}|^{3 + \delta}}  \\
 g(k, \delta, \mathcal{H}) &\equiv& \int_{q \in [k_i, \mathcal{H}]} d^3 q
 \Bigl\{q^2 - \frac{(\,\vect{k} \cdot \,\vect{q})^2}{k^2} \Bigr\}^2 
 \frac{1}{q^{3 + \delta}} \frac{1}{|\,\vect{k} - \,\vect{q}|^{3 + \delta}}  \,,
\end{eqnarray}
where $\delta$ is an arbitrary small constant.
Here we assume
that $k_i$ is much smaller than $k$, i.e., $k_i \ll k$. In fact, $k_i$ is
defined by the horizon scale on the initial time as
$k_i \equiv \mathcal{H}_i$. Since we are interested only in the modes
whose scales are much smaller than the initial 
Hubble horizon scale, it is appropriate to assume $k_i \ll k$.\\
To make the integration simple, we approximate
$|\,\vect{k} - \,\vect{q}|^{-(3 + \delta)}$ as
\begin{eqnarray}
 \mbox{for $q < k$} \hspace{1cm}
 \frac{1}{|\,\vect{k} - \,\vect{q}|^{3+ \delta}}
 &\simeq& \frac{1}{k^{3 + \delta}} \Bigl\{ 1 + 3
 \frac{\,\vect{k} \cdot \,\vect{q}}{k^2} + O \Bigl((q/k)^2 \Bigr)  \Bigr\}
 \nonumber \\
  \mbox{for $k < q$} \hspace{1cm}
 \frac{1}{|\,\vect{k} - \,\vect{q}|^{3+ \delta}} 
 &\simeq& \frac{1}{q^{3 + \delta}} \Bigl\{ 1 + 3
 \frac{\,\vect{k} \cdot \,\vect{q}}{q^2} + O \Bigl((k/q)^2 \Bigr)  \Bigr\}
\,.
 \label{approx for k-q}
\end{eqnarray}
Then $f(k, \delta, \mathcal{H})$ is given by
\begin{eqnarray}
 f(k, \delta, \mathcal{H}) 
 &=& 4 \pi \Bigl[ - \frac{1}{\delta} \frac{1}{k^{3+\delta}}
  \{(k - k_i)^{- \delta} - k_i^{- \delta} \}
  - \frac{1}{3 + 2 \delta} \{ \mathcal{H}^{-3- \delta }
  - (k + k_i)^{-3 - 2 \delta }\} \Bigr] \nonumber \\
 &\simeq& \frac{4 \pi}{k^3} \Bigl[ k^{- 2 \delta} \Bigl\{
 \frac{1}{3} - \frac{1}{\delta} \Bigl( 1 - \Bigl(\frac{k_i}{k} \Bigr)^{- \delta}
 \Bigr)   \Bigr\} - \frac{1}{3} \Bigl( \frac{k}{\mathcal{H}} \Bigr)^3
 \mathcal{H}^{- 2 \delta} \Bigr]
\,.  \label{solution for f}
\end{eqnarray}
Similarly, $g(k, \delta, \mathcal{H})$ is given by
\begin{eqnarray}
 g(k, \delta, \mathcal{H}) &=& \frac{32 \pi}{15}
 \Bigl[ \frac{1}{4 - \delta} \frac{1}{k^{3 + \delta}}
 \Bigl\{ (k - k_i)^{4 - \delta} - k_i^{4-\delta} \Bigr\}
 + \frac{1}{1-2 \delta} \Bigl\{ \mathcal{H}^{1 - 2 \delta}
 - (k + k_i)^{1 - 2 \delta} \Bigr\} \Bigr] \nonumber \\
 &\simeq& \frac{32 \pi}{15} \Bigl( \mathcal{H}^{1 - 2 \delta}
 - \frac{3}{4} k^{1 - 2 \delta} \Bigr)
\,. \label{solution for g}
\end{eqnarray}
\end{widetext}


\bibliographystyle{apsrev} 
\bibliography{sample.bib}

\begin{thebibliography}{99}
\bibitem{footnote2}
If we consider a power-law potential such that $V\propto
\phi^n$, the condition $\varepsilon > \eta_V$ implies $n<2$. 
It means that as the potential becomes flat, the IR divergence will appear.
\end{thebibliography}
%


\end{document}